\documentclass[12pt,twoside]{article}

\usepackage{mathptmx} % This is Times font

\newcommand{\ignore}[1]{}
\usepackage{fancyhdr}
\usepackage[normalem]{ulem}
\usepackage[hyphens]{url}

\usepackage[sort,nocompress]{cite}
\usepackage[final]{microtype}
\usepackage{flushend}

\usepackage{dcolumn}
\usepackage{url}
\usepackage[normalem]{ulem}
\usepackage{amsmath}
\usepackage[usenames,dvipsnames,svgnames,table]{xcolor}
\usepackage{rotating}
\usepackage{enumitem}
\usepackage{booktabs}
\usepackage{multirow}
\usepackage{censor}
\usepackage{listings}
\usepackage{cite}
\usepackage{graphicx}
\usepackage{listings}
\usepackage{color}
\usepackage[final]{pdfpages}
\graphicspath{{./}}
\usepackage{ifthen}
  \newcommand{\rmk}[1]{}
  \newcommand{\rmkdone}[1]{}
  \newcommand{\del}[1]{}
  
  \newcommand{\zdel}[1]{}
  \newcommand{\zadd}[1]{#1}
  \newcommand{\note}[1]{#1}
  
\definecolor{dkgreen}{rgb}{0,0.6,0}
\definecolor{gray}{rgb}{0.5,0.5,0.5}
\definecolor{mauve}{rgb}{0.58,0,0.82}

\usepackage[bookmarks=true,breaklinks=true,letterpaper=true,colorlinks,linkcolor=black,citecolor=blue,urlcolor=black]{hyperref}

\title{{\fontfamily{mono}\fontsize{18pt}{0}\bf\scshape\selectfont BenchIP}: Benchmarking Intelligence Processors}
\date{}

\begin{document}
\maketitle
\vspace{-1.8cm}
\centerline{Jinhua Tao$^1$, Zidong Du$^1$$^,$$^2$, Qi Guo$^1$$^,$$^2$, Huiying Lan$^1$, Lei Zhang$^1$}
\centerline{Shengyuan Zhou$^1$, Lingjie Xu$^3$, Cong Liu$^4$, Haifeng Liu$^5$, Shan Tang$^6$}
\centerline{Allen Rush$^7$,Willian Chen$^7$, Shaoli Liu$^1$$^,$$^2$, Yunji Chen$^1$, Tianshi Chen$^1$$^,$$^2$}
\vspace{0.5cm}
\centerline{$^1$ICT CAS,$^2$Cambricon,$^3$Alibaba Infrastructure Service, Alibaba Group}
\centerline{$^4$IFLYTEK,$^5$JD,$^6$RDA Microelectronics,$^7$AMD}
\vspace{0.2cm}
\begin{abstract}

The increasing attention on deep learning has tremendously spurred the design of intelligence processing hardware. The variety of emerging intelligence processors requires standard benchmarks for fair comparison and system optimization (in both software and hardware). However, existing benchmarks are unsuitable for benchmarking intelligence processors due to their non-diversity and nonrepresentativeness. Also, the lack of a standard benchmarking methodology further exacerbates this problem. In this paper, we propose \textsc{BenchIP}, a benchmark suite and benchmarking methodology for intelligence processors. The benchmark suite in \textsc{BenchIP} consists of two sets of benchmarks: \emph{microbenchmarks} and \emph{macrobenchmarks}. The microbenchmarks consist of single-layer networks. They are mainly designed for bottleneck analysis and system optimization. The macrobenchmarks contain state-of-the-art industrial networks, so as to offer a realistic comparison of different platforms. We also propose a standard benchmarking methodology built upon an industrial software stack and \zdel{hierarchical}\zadd{evaluation} metrics that comprehensively reflect the various characteristics of the evaluated intelligence processors. \textsc{BenchIP} is utilized for evaluating various hardware platforms, including CPUs, GPUs, and accelerators. \textsc{BenchIP} will be open-sourced soon. 

\end{abstract}

\normalsize

\section{Introduction}
\subsection{Motivation}

Recently, \emph{deep learning} has become a \emph{de facto} technique for intelligent processing tasks such as image classification~\cite{Krizhevsky,Simonyan2015,He15CVPR}, video captioning~\cite{Venugopalan15ICCV}, speech recognition~\cite{Ossama14TASLP}, and machine translation~\cite{Akiko16ACL}. As deep learning architectures (i.e., artificial neural networks) are evolving towards deep topologies with complicated transformations, existing general-purpose hardware platforms such as CPUs and GPUs are not able to provide high performance and energy efficiency. As an efficient alternative for deep learning, a large variety of customized hardware architectures, ranging from specialized GPUs (e.g., NVIDIA DGX-1~\cite{nvidiadgx}) to FPGAs (e.g., CNP~\cite{Farabet09FPL} and \cite{Zhang15FPGA}) to ASICs (e.g., DianNao~\cite{Chen14ASPLOS}, NeuFlow~\cite{Farabet2011}, and EIE~\cite{Han2016a}) have emerged. We call those specially designed architectures \emph{intelligence processors} (IPs)\footnote{\pbox[t]{\columnwidth}{Both CPUs and GPUs can be viewed as intelligence processors from a broader perspective, as they remain the mainstream of intelligent processing.}}. 

Benchmarking has served long as the foundation of designing new hardware architectures. For example, the SPEC-CPU series~\cite{speccpu} and the corresponding methodology (e.g., framework, procedure, and metrics, etc.) have been the incontestable source for evaluating and optimizing general-purpose architecture, and they have evolved over time to keep pace with the rapid advance of unicore architecture. With multicore architectures dominating the market, the PARSEC benchmark suite~\cite{parsec} was released in 2008 accordingly. \zadd{Recently, especially in last one or two years, intelligence processors have become a hot topic. However, with the chaos of benchmarks, it is difficult to compare and quantize the various IPs, especially the rapidly increasing works emerging at top architecture conferences~\cite{Han2016a, Alwani2016,Judd2016,Rhu2016,Zhang2016a,Ji2016,Kim2016a,LiKamWa2016,Albericio2016,Chi2016,Shafiee2016,Liu2016a,Chen2016b,Reagen2016,Song2017,Lu2017,Wzr2017}. Still, fair and comprehensive benchmarking is a necessity for assessing the progress of IPs.}\zdel{Recently, as intelligence processors are becoming a hot topic in top architecture conferences~\mbox{\cite{Han2016a,Shafiee2016,Chen2016b,Liu2016a,Chi2016}}, fair and comprehensive benchmarking is a necessity for assessing the progress of IPs.} Moreover, with the help of benchmarking, both architects and practitioners can compare various architectures, identify their bottlenecks, and conduct corresponding system/architectural optimization. As a result, during the design and deployment of intelligence processors, one of the central tasks is benchmarking.

\textbf{Benchmark suite.} When benchmarking intelligence processors, designing an appropriate benchmark suite is the top-most consideration. Roughly, two main categories of benchmarks already exist: the \emph{collective} benchmarks, which are collected and documented as a benchmark suite (e.g., BenchNN~\cite{Chen2012a}, DeepBench~\cite{Baidu16DeepBench}, \zadd{and Fathom~\cite{Adolf2016}}), and the \emph{personalized} benchmarks, which have been used to evaluate different intelligence processors in various computer architecture papers.

BenchNN illustrates the potential of neural networks by using them to re-implement five tasks from the well-known PARSEC benchmark suite to reveal the broad application scope of neural networks. Though BenchNN successfully demonstrated an urgent need for neural networks, it is more like a symbol rather than a practical benchmark suite. To be specific, there are three limitations: (1) BenchNN provides a very limited number of neural network applications (i.e., five in total), (2) the neural networks used in BenchNN are classic models such as simple multi-layer perceptrons that fails to characterize the latest trend reflected by state-of-the-art deep learning techniques, and (3) BenchNN does not reflect the significant diversity among neural networks---for example, both the \emph{ferret} and \emph{streamcluster} benchmarks are built on the Self Organizing Maps~\cite{Murtagh95JoC}.

DeepBench aims at measuring the performance of basic deep learning operations across various hardware platforms. Such operations include matrix multiplication, convolution, recurrent layer, and all-reduce operations. As DeepBench only contains basic primitive operations, it cannot approximate high-level behaviors of the full-stack neural networks widely used in industry. For example, the Faster-RCNN~\cite{Ren} used for object detection contains far more operations than the basic operations in DeepBench. In short, due to its inability to evaluate of the full-stack neural networks used in practice, DeepBench is not suitable for benchmarking intelligence processors.

\zadd{Fathom focuses on better understanding a broader variety of deep learning workloads. However, with an assembled set of only eight archetype deep learning workloads, Fathom has limited capability for benchmarking intelligence processors. First, Fathom fails on diversity, as it only contains common layers such as~\emph{Conv.} and \emph{Fully-Connected} but lacks important layers such as~\emph{Deconv., Unpooling} and \emph{Batch normalization}. Second, Fathom fails to provide a benchmarking method for several of the most important categories of intelligence processors, e.g., customized hardware architectures. Third, Fathom cannot benchmark fine-grained performance or efficiency, which are important for intelligence processors, especially customized hardware. Thus, Fathom is also not suitable for benchmarking intelligence processors.}

\begin{figure}[t]
  \centering
  \vspace{-0.7cm}
  \includegraphics[width=\columnwidth]{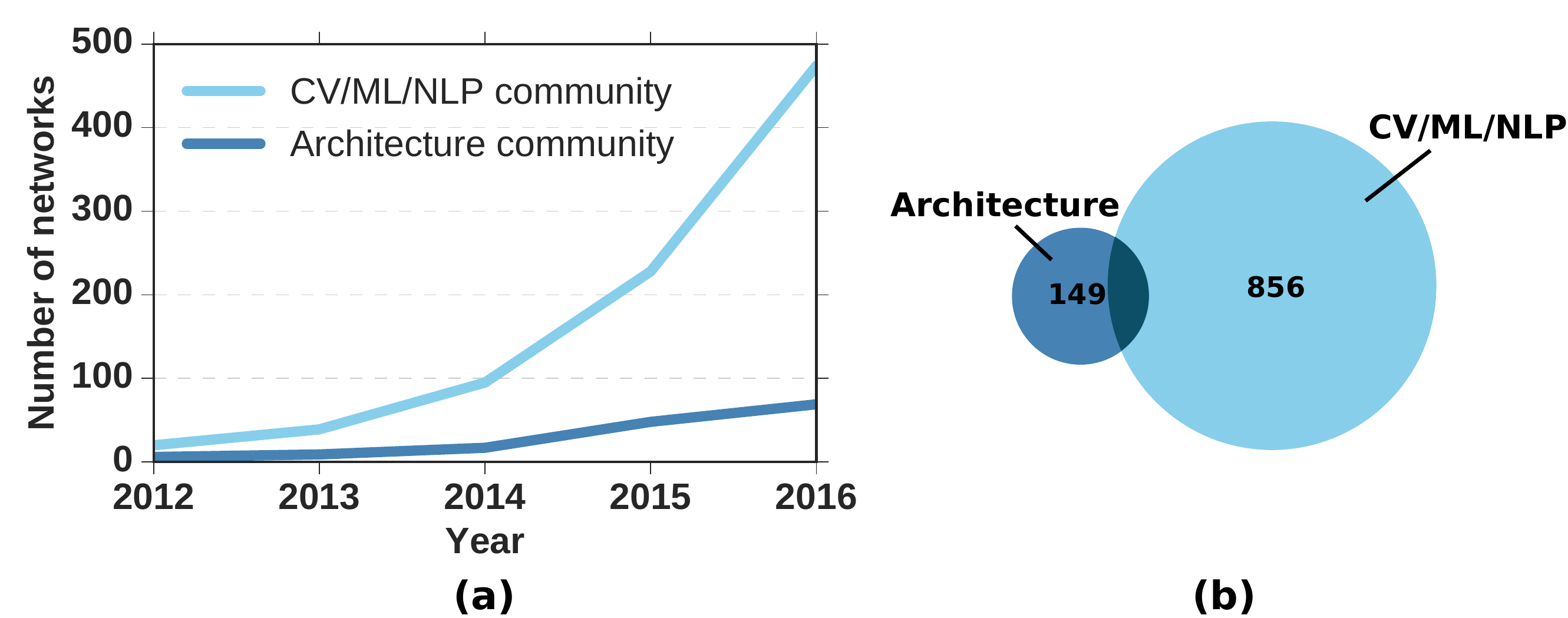}
    \vspace{-0.86cm}
  \caption{\footnotesize \emph{(a)}: The number of neural network models proposed in CV/ML/NLP top-tier conferences (e.g., CVPR/ICML/EMNLP) has increased rapidly, while the number of neural networks utilized in the computer architecture community (e.g., ISCA/MICRO/HPCA/ASPLOS) exhibits relatively moderate growth.~\emph{(b)}: The neural networks used in the computer architecture community only cover a small range of the neural networks used in applications from the AI community.}
  \label{fig:gap}
\vspace{-0.4cm}
\end{figure}

The \emph{personalized} benchmarks are the neural networks that appear in computer architecture papers \zadd{but not as a benchmark suite} for evaluating intelligence processors. Such benchmarks include both basic operations (e.g., convolution, activation, and normalization) and full neural networks (e.g., AlexNet~\cite{Krizhevsky}, VGG~\cite{Simonyan2015}, and ResNet~\cite{He15CVPR}). As such benchmarks vary for evaluating different architectures, they may suffer from non-diversity and nonrepresentativeness. Figure~\ref{fig:gap}(a) shows that there exists a large gap between the neural networks proposed in CV (computer vision)/ML(machine learning)/NLP (natural language processing) applications and those used for evaluating hardware architectures. \zadd{We also notice the increasing capability of performing multiple types of neural networks in the architecture community.} Interestingly, from Figure~\ref{fig:gap} (b), we observe that more than half of the personalized benchmarks utilized in computer architecture papers are \emph{not} practically used in AI research fields (e.g., CV/ML/NLP). Even when we assume a one-year delay for transferring knowledge from the AI community to the computer architecture community, this number is still larger than 30\%. This observation clearly shows a risk that conclusions made by the computer architecture community might mislead the AI community.

In short, due to their non-diversity and nonrepresentativeness, existing benchmarks are not suitable for benchmarking intelligence processors.

\textbf{Benchmarking methodology.} A benchmark suite only specifies \emph{what} to measure. We also need to determine \emph{how} to benchmark intelligence processors, i.e., develop a benchmarking methodology. The existing benchmarking methodology utilized in the computer architecture community is chaotic, mainly due to the complicated software/hardware stacks of modern computer systems. Given the prototypical nature of a newly published intelligence architecture, a full software stack may not exist to conduct a fair comparison against commodity hardware. As a result, some evaluate their architectures with a high-level programming framework~\cite{Chen2016b,Albericio2016}, while others are evaluated directly at the RTL level by getting rid of the software stack~\cite{Han2016a}. Since the software stack in real applications is very crucial to overall performance, comparison at different hardware/software levels seems to be a case of comparing ``apples and oranges." To alleviate this problem, a benchmark specification for the evaluation framework, metrics, and procedure should be clearly defined.

Moreover, intelligence processors have unique features compared with general-purpose architectures that pose further challenges for the methodology. One such feature is that the prediction accuracy is also an important design tradeoff in addition to performance/power/area. While it could be a reasonable solution to slightly relax the prediction accuracy of machines learning models (via adopting, e.g., data quantization~\cite{Rastegari2016a} or connection sparsity~\cite{Song2015}) for performance and energy efficiency, excessive relaxation on the prediction accuracy will lead to unfair and degenerate comparisons. Therefore, it is necessary to take the accuracy, together with other traditional design tradeoffs, into consideration when designing the benchmarking methodology.

\subsection{Our work}

Based on the above observations, in this paper, we propose \textsc{BenchIP}, a novel benchmark suite and benchmarking methodology deliberately designed for intelligence processors. The benchmark suite in \textsc{BenchIP} contains two types of benchmarks: \emph{microbenchmarks} and \emph{macrobenchmarks}. The microbenchmarks contain 12 representative single-layer networks, such as convolution, pooling, activation, etc., that are mainly used for architectural/system optimization. The macrobenchmarks consist of 11 commonly-used full neural networks (e.g., AlexNet, VGG, and Faster-RCNN) from different intelligent processing scenarios. These are directly extracted from real applications and they are mainly used for the evaluation and comparison of hardware platforms.

Our benchmarking methodology is mainly designed for creditability, portability, and fairness. The creditability is achieved by building an industrial software stack consisting of a high-level programming model, library, and device driver. The portability is accomplished by providing a standard interface for the intermediate-level high-performance library. Thus, with a new architecture, once the corresponding library complies with the standard interface, it can be easily plugged into the software stack for evaluation. For the fairness purpose, we \zdel{propose hierarchical evaluation}\zadd{report evaluation} metrics that take the accuracy into consideration in order to comprehensively reflect the various characteristics of intelligence processors.

\textbf{Contributions.}  The work in this paper makes the following key contributions:

\begin{itemize}[leftmargin=0.5cm,topsep=0cm,itemsep=0.0cm]
\item We build a benchmark suite consisting of microbenchmarks and macrobenchmarks to evaluate the accuracy, performance, and energy of intelligence processors.
  \vspace{-0.1cm}
\item We conduct a comprehensive analysis of the proposed benchmarks in order to demonstrate that the proposed benchmark suite is representative and diverse.
    \vspace{-0.1cm}
  \item We propose a benchmarking methodology that contains an industrial-level software stack and \zdel{hierarchical}\zadd{evaluation} metrics for guaranteeing creditability, portability, and fairness.
      \vspace{-0.1cm}
\item We evaluate various intelligence processors, including CPUs, GPUs, and accelerators, with \textsc{BenchIP}.
\end{itemize}

\begin{table*}[!t]
  \centering
  \scriptsize
  \caption{The benchmarks in \textsc{BenchIP}}
  \label{tab:benchip}
  \begin{minipage}[c]{\textwidth}
    \centering
  \begin{tabular}{lllll||lllll}
    \toprule
    Layer &&\# Config. & Note && NN & Dataset & Scenario \\
    \midrule
    Conv. & & 7 (3/1/3)\footnotemark[1] &&& LeNet-5~\cite{YannLeCun1998} & MNIST~\cite{YannLeCun1998}& Hand written digits recognition &\\
    Pooling & &7/7 (3/1/3) & Avg./Max && RNN~\cite{Graves2014} & WSJ~\cite{Marcus1993}&Speech recognition &\\
    FC & & 7 (3/1/3) &   && AlexNet~\cite{Krizhevsky} & ImageNet~\cite{Russakovsky2014}&Image classification&\\
    ReLU & & 7 (3/1/3) & && VGG~\cite{Simonyan2015} & ImageNet~\cite{Russakovsky2014}&Image classification&\\
    Sigmoid &  & 7 (3/1/3)& && ResNet~\cite{He} & ImageNet~\cite{Russakovsky2014}&Image classification&\\
    LRN & & 7 (3/1/3)& && Faster R-CNN~\cite{Ren} &PASCAL VOC 2012~\cite{pascal-voc-2012}& Object recognition&\\
    BN & & 7 (3/1/3)& && Deep Face Recog.~\cite{Parkhi2015} & LFW~\cite{LFWTech} &Face recognition& \\
    Unpooling & & 7/7 (3/1/3)& Avg./Max && DeconvNet~\cite{Noh2015} & PASCAL VOC 2012~\cite{pascal-voc-2012}&Semantic segmentation&\\
    Deconv. & & 7 (3/1/3)&  && FCLN~\cite{Johnson2015} &Visual Genome~\cite{Johnson2015}&Image captioning&\\
    LSTM & & 7 (3/1/3) & && S2VT~\cite{Venugopalan} &MSVD~\cite{Chen2011}&Video captioning&\\
     & & & && SyntaxNet~\cite{Andor2016} &English WSJ~\cite{Marcus1993}&Nature language processing&\\
    \bottomrule
  \end{tabular}
  \protect{\footnotetext{\scriptsize $^1$Total configurations (Normal configurations/Extreme small configurations/Extreme large configurations)}}
  \end{minipage}
\end{table*}

\vspace{-0.2cm}
\section{The Benchmark Suite}
\label{sec:benchmark}
\vspace{-0.1cm}

In this section, we first introduce the requirements of an ideal benchmark suite for intelligence processors. We next present our benchmarks in detail. Finally, we conduct an analysis to demonstrate the rationality of the proposed benchmarks.

\vspace{-0.2cm}
\subsection{Design requirements}
\label{subsec:requirement}
An ideal benchmark suite for intelligence processors should meet the following requirements from both the application and architecture perspectives.

\begin{itemize}[leftmargin=0.5cm,topsep=0cm,itemsep=0.0cm]
\item \textbf{Application perspective.}  The selected benchmarks should include mainstream neural network algorithms from a broad range of application scenarios. Moreover, in order to reduce benchmarking efforts, benchmarks with similar characteristics should not be included redundantly~\cite{Phansalkar07ISCA}.
\item \textbf{Architecture perspective.} In addition to the common usage scenarios, the selected benchmarks should be able to explore the processing boundary of underlying architectures, as well as address the future trends for intelligence processors.
\end{itemize}

\vspace{-0.1cm}
\subsection{Benchmarks}
\vspace{-0.1cm}

Based on the above requirements, we construct both \emph{microbenchmarks} and \emph{macrobenchmarks} in \textsc{BenchIP}, particularly for system optimization and platform comparison, respectively. In the context of deep learning, the microbenchmarks only include single-layer networks, while the macrobenchmarks contain entire neural networks extracted from real applications. For the microbenchmarks, individual layers are equipped with multiple configurations, including normal and extreme cases for stress-testing intelligence processors. For the macrobenchmarks, entire neural networks are used for processing end-to-end intelligent tasks where data movements between layers are also crucial.

\rmk{Reviewer A: present in alphabet order?}

\vspace{-0.2cm}
\subsubsection{Microbenchmarks}
Microbenchmarks are widely used for system and architectural optimization in both industry and academia. These benchmarks measure a specific component of the system. For example, STREAM is a well-known microbenchmark for evaluating the memory bandwidth of the system~\cite{McCalpin95TCCA}. The University of Edinburgh provides the OpenMP microbenchmark suite for measuring the overheads of synchronization, loop scheduling, and array operations in the OpenMP runtime library~\cite{Bull01CAN}.

In the context of deep learning, the microbenchmarks refer to single-layer networks. \zadd{Unlike DeepBench, we do not select basic operations as microbenchmarks because basic operations cannot benchmark the efficiency of data movements}. We include 10 single-layer networks, as listed in Table~\ref{tab:benchip}. In order to fulfill the requirements mentioned in Section~\ref{subsec:requirement}, the principle is threefold: the selected layers should be widely used in existing neural networks (e.g.,~\emph{convolutional layer}); such layers should have significant different computational, memory, and control patterns; and such layers may have huge impacts on future designs (e.g.,~\emph{deconvolution layer}).

\textbf{Conv.} The convolutional layer is the most important layer in deep learning, especially for images/video tasks as it automatically extracts features from the input images/videos (2D/3D input data) by applying a set of 2D/3D filters.\zdel{In a convolutional layer, the 2D output feature maps are obtained at last where each output feature map is the convolutional results between a set of filters in same size (i.e., usually called kernel) and neurons at the same location in a set of input feature maps. In particular, for a output feature map, the kernels can be \emph{private} for each neuron or \emph{shared} among all the neurons.} Regarding \emph{representativeness}, we study network layers employed in application scenarios from the AI community (e.g., the CV/NLP/ML community). More specifically, we count the appearance time of a layer/network (\# Presence) in the papers of such fields and thus the appearance ratio to the total layers/networks (\# Ratio). These results are shown in Table~\ref{tab:software_statistics}. From the Table, we can see that convolutional layer appears in 19.28\% of all layers.

\textbf{Pooling.} In deep learning, the pooling layer is required to aggregate features for later classification by downsampling 2D input data.\zdel{ The output neurons are calculated by directly downsampling the input feature maps with a maximum or average operation in local non-overlapped/overlapped windows of the input feature maps.} As shown in Table~\ref{tab:software_statistics}, the pooling layer is as important as the convolutional layer, having a 18.76\% appearance ratio among all layers.

\begin{table}[!t]
\vspace{-0.2cm}
  \centering
  \scriptsize
  \caption{Statistics of the different types of layers in CVPR, EMNLP, ICML, ICRA, and NIPS (2012$\sim$2016).}
  \label{tab:software_statistics}
  \begin{minipage}[c]{\columnwidth}
  \begin{tabular}{p{1.4cm}@{}p{1.3cm}@{}p{1.3cm}|p{1.3cm}@{}p{1.3cm}@{}p{1cm}}
    \toprule
    Layer Type & \# Presence & Ratio(\%) & Network & \# Presence & Ratio(\%)\\
    \midrule
    Conv. & 452 & 19.28 & LeNet & 17  & 1.99 \\
    Pooling\footnotemark[1] & 440 & 18.76 & AlexNet & 159 & 18.57 \\
    FC & 473 & 20.17 & VGG & 171 & 19.98 \\
    LRN & 181 & 7.72 & ResNet & 5 & 0.58 \\
    ReLU & 425 & 18.12 & Faster-RCNN & 34 & 3.97 \\
    Sigmoid & 90 & 3.84 & CNN & 128 & 14.95 \\
    Tanh & 100 & 4.26 & DNN & 82 & 9.58 \\
    Deconv. & 19 & 0.81 & RNN & 260 & 30.37 \\
    Unpooling & 16 & 0.68 & Others\footnotemark[2] & 30 & 1.28 \\
    LSTM & 133 & 5.67 & & & \\
    BN & 16 & 0.68 & & & \\
    \midrule
    Total & 2345 & - & Total & 856 & -  \\
    \bottomrule
  \end{tabular}
  \protect{\footnotetext{\scriptsize $^1$Average Pooling + Max Pooling}
    \footnotetext{\scriptsize $^2$RBM, SNN, and other ML techniques}}
  \end{minipage}
\end{table}

%\noindent
\textbf{FC.} The fully-connected (FC) layer is the most common layer, where each output neuron is connected to all input neurons with independent synapses.\zdel{ Output neuron $y_{i}$ is computed with the expression of $y_{i}=f(\beta+\sum_{j}\omega_{i,j} x_{j})$, where $\omega_{i,j}$ is the synaptic weights between input neuron $x_{j}$ and output neuron $y_{i}$, $f(\cdot)$ is the activation function and $\beta$ is the bias.} According to Table~\ref{tab:software_statistics}, it has been employed in most application scenarios, with an appearance ratio as high as 20.17\%.

%\noindent
\textbf{ReLU.} Nonlinear activation layer is a key component as it introduces nonlinearity into the existing deep neural networks to improve the capability of classification. The inputs and outputs can be treated as one-dimensional data with element-wise operations. Rectified Linear Units (ReLU)\zdel{, which applies non-saturating nonlinear function $f(x)=max(0,x)$ to the input neurons,} is becoming popular (with an appearance ratio of 18.12\%) as it achieves better accuracy.

%\noindent
\textbf{Sigmoid.} The Sigmoid layer is another type of activation layer. \zdel{It applies nonlinear function $f(x) = \frac{1}{1+{e^{-x}}}$ to each input neuron. }While the Sigmoid layer is becoming less popular, the appearance ratio is still about 3.84\% of all employed network layers. \zadd{Both ReLU and Sigmoid are typical element-wise operations but have very different functionality. Thus, we include both as microbenchmarks.}

\textbf{LRN.} The normalization layer introduces competition between input neurons at the same position on different input feature maps, which have significantly different computation patterns from other layers. There exists two typical normalization layers: Local Response Normalization (LRN)~\cite{Krizhevsky} and Local Contrast Normalization (LCN)~\cite{Du2015}. We include LRN as the representative layer for its relative high appearance ratio (7.72\%) compared to many other layers (including LCN). In an LRN layer, output neurons are the results of input neurons divided by responses among neighboring feature maps in the same spatial location.

%\noindent
\textbf{BN.} Batch normalization performs normalization to each mini-batch, which allows networks to use larger learning rate and thus to accelerate neural network convergence in training. It also helps reduce the dependency on weight initialization.\zdel{ During inference, the input is first normalized, then transformed through a linear function $f(x) = \gamma\cdot x + \beta$, where $\gamma$ and $\beta$ are learned parameters.} We select it as a microbenchmark due to the increasing trend of using BN to accelerate the training.

\textbf{Deconv.} The deconvolutional layer is emerging in the fields of feature extraction, image reconstruction, and image semantic segmentation as it increases the size of inputs for better matching and locating. It can be viewed as a reverse operation of convolution\zdel{, where the output features are mapped back to the input space}. We include it as a microbenchmark for its unique computational pattern and the increasing popularity of employing such a layer in emerging applications (e.g., semantic segmentation).

\textbf{Unpooling.} The unpooling layer has a purpose and effect similar to the deconv layer. It performs the reverse operation of pooling and reconstructs a larger size of activations. \zdel{The inputs are mapped to corresponding positions of the reconstructed output according to locations of activated neurons. }The unpooling operations have gained increasing attentions in recent years, as semantic segmentation and image reconstruction have become more popular.

\textbf{LSTM.} The LSTM layer is a variant of the recurrent layer, which prevents the gradient from vanishing or exploding. It contains multiple element-wise gate operations and active operations (i.e., Sigmoid and Tanh). It is selected for its high appearance ratio of 5.67\% and also because of the representative of recurrent computing patterns.

\textbf{Summary.} In Table~\ref{tab:software_statistics}, almost all appearing layers are included as microbenchmarks except for Tanh, which is mostly used as a part of the LSTM layer. Thus, it is fair enough to say that our chosen microbenchmarks for \textsc{BenchIP} has been thorough and rational.

\vspace{-0.3cm}
\subsubsection{Macrobenchmarks}
Macrobenchmarks indicate workloads that are directly extracted from real industrial applications for evaluating and comparing different platforms---for example, in SPEC CPU-2006~\cite{speccpu}, all benchmarks are collected and documented from desktop software for approximating real application scenarios. The PARSEC benchmark suite comprises emerging workloads from various areas, e.g., computer vision, financial analytics, and animation physics~\cite{parsec} and is aimed at benchmarking dominant chip multiprocessors.

In the context of deep learning, macrobenchmarks refer to entire neural networks such as AlexNet, VGG, and Faster-RCNN. We selected the 11 entire networks, as listed in Table~\ref{tab:benchip}. There are three reasons for having entire networks as macrobenchmarks: (1) compared with single-layer networks in microbenchmarks, entire networks are commonly used for performing end-to-end tasks such as speech recognition; (2) using combinations of individual layers for evaluation is not sufficient because factors such as the data exchange when coupling two layers are critical to the performance and energy efficiency, especially as networks become deeper and larger with increasing processing complexity; and (3) the optimization between layers cannot be measured with microbenchmarks, which are only capable of evaluating \zdel{inner-layer}\zadd{inter-layer} optimizations.

\textbf{LeNet-5.} LeNet-5 is a notable \zadd{convolutional }neural network used to classify handwritten and machine-printed characters.\zdel{ The network consists of two convolutional layers, each followed by a pooling layer, and three fully-connected layers.} It is selected as the representation of handwritten digits recognition.

\textbf{RNN.} RNN is a network for end-to-end speech recognition. It consists of a deep bidirectional LSTM recurrent neural network and a Connectionist Temporal Classification (CTC) objective function. We select it to represent the speech recognition scenario.

\textbf{AlexNet.} AlexNet is a deep convolutional neural network designed for the classification task of ImageNet. \zdel{The network is composed of five convolutional layers, some of which followed by max-pooling layers, two LRN layers, and three fully-connected layers. }We select it as it is a very popular and widely used network \zadd{in both architecture and AI communities}, with appearance ratios of 33.54\% \zadd{and 18.57\%, respectively (See Table 2)}. 

\textbf{VGG.} VGG is a series of networks with very deep architecture designed for image classification. It uses very small kernels (e.g., $3 \times 3$ and $1 \times 1$) for convolutional layers through the whole network. It is also a very popular network with the appearance ratio of 36.08\% \zadd{in architecture community and 19.98\% in AI community}.

\textbf{ResNet.} ResNet refers to the deep residual networks. The residual learning framework allows it to achieve extremely deep architecture, and it gains high accuracy from the increased depth. It uses ``shortcut connections" to skip one or more layers to ease the degradation problem. We select ResNet because it is the state-of-the-art network for image classification.

\textbf{Faster R-CNN.} Faster R-CNN is an object detection system with high performance and detection accuracy. It is composed of two modules: a region proposal network (RPN) for generating region proposals, and a fast R-CNN detection network that uses such proposals for detection. We select it as the representation of object detection network.

\textbf{Deep face recognition.} Deep face recognition utilizes deep CNN networks to classify and verify face images. The networks are derived from the \zdel{VGG11, VGG13, VGG16 }\zadd{VGG }architectures with the last FC layer replaced by a classifier or an embedding descriptor vector. We select it as the representation of face recognition scenario.

\textbf{DeconvNet.} DeconvNet is a deconvolutional network for semantic segmentation that consists of two subnetworks: convolution and deconvolution.\zdel{The convolution network, which is based on the VGG16 architecture, extracts features from the input images and generates a multidimensional feature representation. The deconvolution network takes the representation as inputs and produces object segmentations.} We choose this network due to the deconvolutional and unpooling layers it contains, and as the representation of semantic segmentation scenario.

\textbf{FCLN.} FCLN is a learning system for image caption tasks, which is composed of a convolution network and a recurrent network. The input images are first processed by a VGG16-based convolution network and then fed into the localization layer to identify regions of interest. These regions are input into the recurrent language model to generate captions. We selected FCLN as the representative network of image caption scenario.

\textbf{S2VT.} S2VT is a sequence-to-sequence LSTM model trained on video-sentence pairs for generating descriptive natural language text of video clips. It uses a VGG16-based network to extract features from the raw image and then feeds the features into the LSTM network, which decodes the representation into a sequence of words. We selected S2VT as the representation of video caption scenario.

\textbf{SyntaxNet.} SyntaxNet is a globally normalized transition-based feed-forward network for part-of-speech tagging, dependency parsing, and sentence compression. It at core is a transition-based parser, where neural network is adopted for computing the score of decisions in certain states. We select it to represent the natural language processing scenario.

\textbf{Summary.} Macrobenchmarks include well-known neural networks, e.g,~\emph{LeNet-5, AlexNet} and~\emph{VGG}, as well as typical networks such as~\emph{RNN}. Thus,~\textsc{BenchIP} has both representativeness and diversity as it well covers 98.72\% of the networks that appeared in application domains in the last five years. Especially, note that most authors of this paper are from both hardware and AI industry/academia, and the selected benchmarks are commonly used in our daily production environments.

\vspace{-0.3cm}
\subsection{Benchmark analysis}
\label{subsec:benchip_analysis}
In this section, we conduct a comprehensive analysis of the selected benchmarks to demonstrate that they can meet the requirements from both the application and architecture perspectives. The basic intuition is to characterize the benchmarks using architecture-independent characteristics. As the macrobenchmarks are composed of single-layer networks---for example, the microbenchmark \emph{Conv.} layer provides a configuration that is exactly the second layer in the macrobenchmark \emph{VGG}---we focus on analyzing the microbenchmarks.

\begin{figure}[!t]
  \vspace{-0.5cm}
  \begin{minipage}[b]{0.49\columnwidth}
    \centering
    \hspace*{-0.5cm}
    \includegraphics[width=1.1\columnwidth]{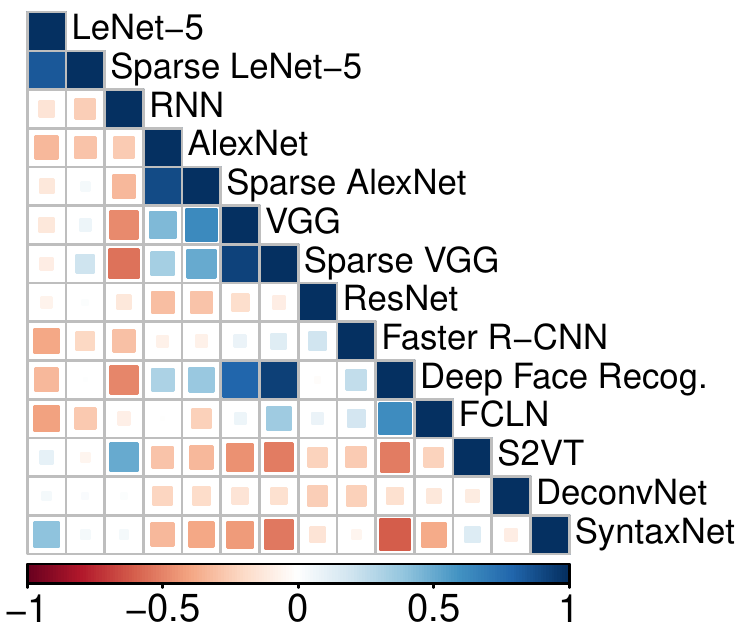}
\end{minipage}
\hfill
  \begin{minipage}[b]{0.49\columnwidth}
    \centering
    \hspace*{-0.3cm}
    \includegraphics[width=1.15\columnwidth,height=3.8cm]{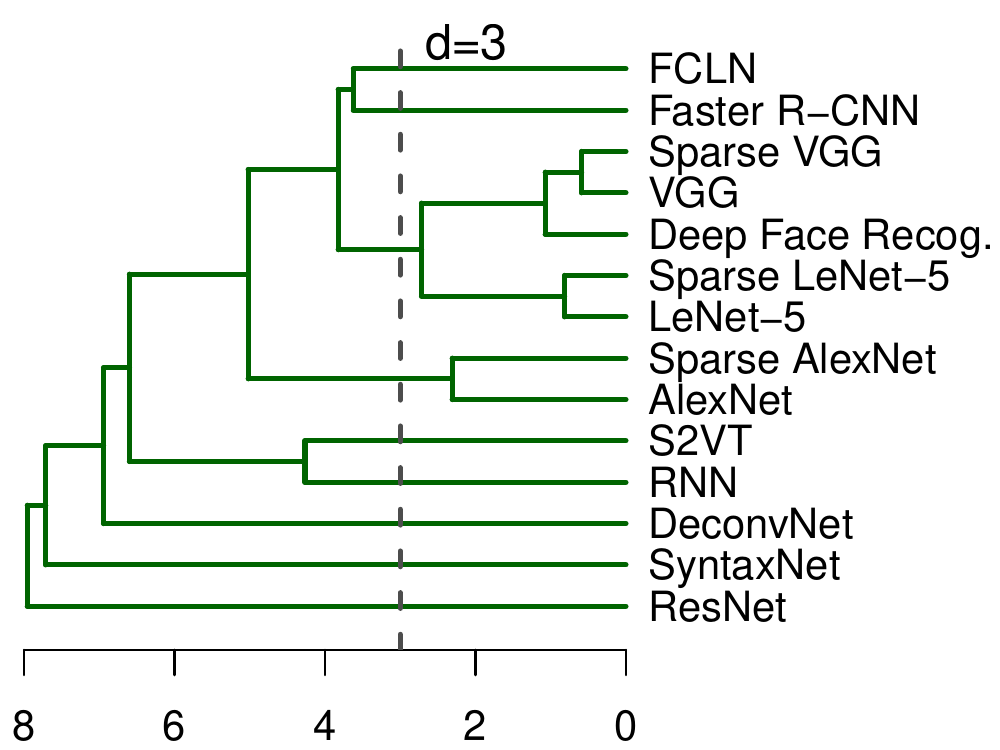}
    \end{minipage}
  \vspace{-0.3cm}
  \caption{\emph{Left:} Correlation heatmap of macrobenchmarks.~\emph{Right:} Similarity of macrobenchmarks.}
  \label{fig:macrobench}
    \vspace{-0.7cm}
\end{figure}

\zadd{In Figure~\ref{fig:macrobench}, we still report the similarity of macrobenchmarks to better understand the variation of macrobenchmarks. We use operation amounts in different layers as feature vectors and Euclidean distances to hierarchically cluster similar networks into groups (see right part of Figure~\ref{fig:macrobench}). The x-axis is the linkage distance between two macrobenchmarks or two clustered groups; similar networks will be grouped first. Similarly, we report the correlation of macrobenchmarks using a heatmap as shown in the left part of Figure~\ref{fig:macrobench}. Each square block in various sizes or colors shows the correlation between the two networks which are indicated by the top and right labels. It is not surprising to observe the close relationship between the sparse and dense versions of a network as they share the same network architecture but have a different amount of operations. For example,~\emph{sparse VGG} and~\emph{VGG} merge early in the dendrogram and have cooler colors (more related) in heatmap. In addition, half the macrobenchmarks have distances larger than 3.12 (geometric mean distance, 39\% of the longest distance). As well, the heatmap is almost occupied by square blocks with warmer color (less related). It clearly shows the diversity of the macrobenchmarks achieved by carefully considering the representativeness and diversity from the perspective of application and architecture (see Section~\ref{subsec:requirement}).} 

\vspace{-0.3cm}
\subsubsection{Characteristics}
The employed architectural-independent characteristics are listed in Table~\ref{tab:char}. They can be classified into three categories: \emph{memory}, \emph{computation}, and \emph{control}, which are the main factors of overall efficiency.

\textbf{Memory characteristics.} As the memory wall continues to grow, we first consider three types of memory characteristics: the number of memory accesses (MemAcc), reuse distance (ReDist), and the memory footprint. The number of memory accesses is a direct metric of memory-intensiveness. Reuse distance refers to the number of different data elements accessed between two consecutive reuses of the same element~\cite{Ding03PLDI}. It has long been used for measuring locality behaviors. Memory footprint refers to the amount of memory accessed at runtime. It is crucial to systems, which are sensitive to memory capacity such as embedded systems. \zadd{In the context of neural networks}\zdel{For our microbenchmarks}, memory footprint can be further classified into three groups: input (InMem), output (OutMem), and weight (WghMem).

\begin{table}[!t]
\vspace{0.1cm}
  \centering
  \scriptsize
  \caption{Architecture-independent characteristics}
  \label{tab:char}
  \begin{minipage}[c]{\columnwidth}
  \begin{tabular}{lllllllllll}
    \toprule
    Category & Name  & Notes \\
    \midrule
    Memory & MemAcc & the number of total memory accesses\\
           & ReDist & reuse distance\\
           & InMem & memory size of input\\
           & OutMem & memory size of output\\
           & WghMem & memory size of weight\\
    Computation & Ops& the number of operations\\
                & OpMem & the ratio of operations to memory access\\
                & ComPtt & computation patterns\\
    Control & PR & branch prediction ratio\\
            & MPR & misprediction ratio \\
    \bottomrule
  \end{tabular}
  \end{minipage}
\end{table}

\rmk{Reviewer A: what are the computations?}
\textbf{Computation characteristics.} We define three types of computation characteristics: the number of operations (Ops), the ratio of operations to memory access (OpMem), \zadd{and the computation patterns (ComPtt)}. The number of operations is a direct indicator of the problem size. The ratio of operations to memory access can determine whether an algorithm is compute- or memory-intensive. Also, regarding the computation patterns of deep learning algorithms, there are three types: reduction (RD), element-wise (EW) and enlargement (EL). RD refers to operations that transform multiple input neurons into a single neuron, e.g., \emph{Conv.}, \emph{Pooling}, \emph{FC}, and \emph{LRN}. EW refers to operations that enforce element-wise transformations on the input neurons, e.g., \emph{ReLU}, \emph{Sigmoid}, and \emph{BN}. EL is the operation that transforms one or a set of input neurons into multiple output neurons, e.g., \emph{Deconv.} and \emph{Unpooling}.

\textbf{Control characteristics.} The control characteristics are closely related to branch behaviors. We use two metrics to measure control characteristics: branch prediction ratio (PR) and the misprediction ratio (MPR). The branch ratio is the number of branches to that of all instructions, and the misprediction ratio is the number of mispredicted branches to the number of total branches.

\begin{figure}[!t]
    \vspace{-0.4cm}
  \centering
    \hspace*{-0.6cm}
    \includegraphics[width=1.1\columnwidth,height=18cm]{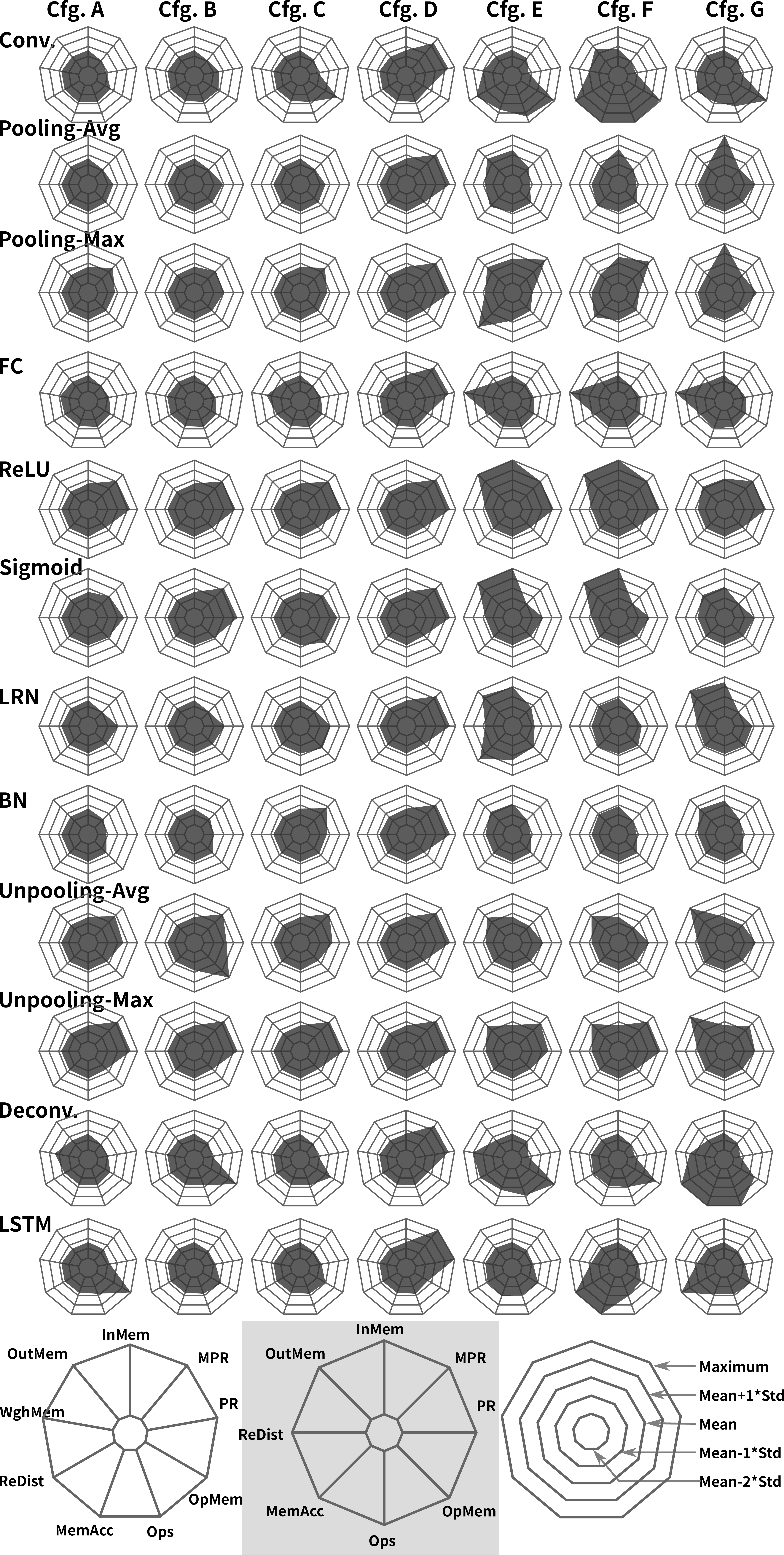}
    \vspace{-0.6cm}
  \caption{Kiviat chart of all configurations of microbenchmarks.}
  \label{fig:hex}
\vspace{-0.4cm}
\end{figure}

\subsubsection{Application perspective}
\vspace{-0.07cm}

Though our benchmarks include most representative neural networks from a wide range of application scenarios, their inherent characteristics should be diverse to reduce the redundancy of the benchmark suite. Based on the aforementioned architecture-independent characteristics, the diversity can be quantitatively measured. Intuitively, there are two types of diversities in the \textsc{BenchIP} benchmark suite: inter-layer and intra-layer diversity. The inter-layer diversity is achieved with 12 different benchmarks, and the intra-layer diversity is achieved with seven configurations per benchmark, including three normal configurations (Cfg.\,A$\sim$\,C), one small configuration (Cfg.\,D), and three large (extreme) configurations (Cfg.\,E$\sim$\,G). The normal configurations are directly extracted from commonly used entire networks, while the extreme configurations can be used for stress testing hardware architectures.

\begin{figure}[!t]
    \vspace{-0.5cm}
  \centering
      \hspace*{-0.6cm}
      \includegraphics[width=1.08\columnwidth]{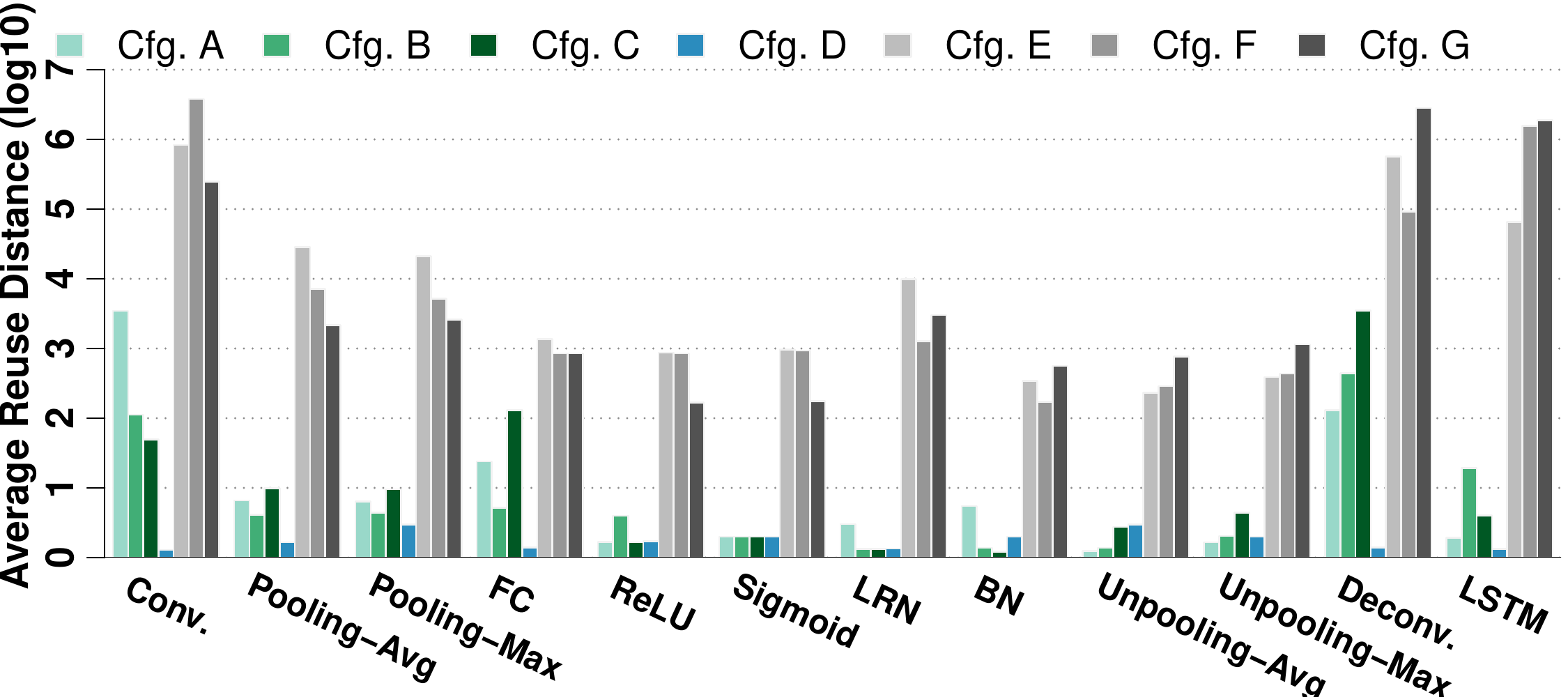}
    \vspace{-0.8cm}
  \caption{Average reuse distances of microbenchmarks.}
  \label{fig:avg_reuse_distance}
\vspace{-5pt}
\end{figure}

Figure~\ref{fig:hex} shows the kiviat chart of seven configurations from all 12 microbenchmarks. The axes represent characteristics listed in Table~\ref{tab:char}\footnote{The ComPtt characteristic is not included in this figure since there are only three categories.}, and the meaning of each ring is shown in Figure~\ref{fig:hex} as well. We can observe that the characteristics of the extreme configurations are significantly different from those of the normal configurations, which well demonstrates the intra-layer diversity. For instance, for the \emph{FC} layer, the weight size of extreme configurations (e.g., Cfg.\,F) is much larger than that of the normal configurations (e.g., Cfg.\,C). Moreover, even for the extreme configurations, as they are carefully designed to emphasize different characteristics, the diversity between them is also obvious, e.g., the input size of Cfg.\,F is 5.3 times smaller than that of Cfg.\,G for the \emph{Pooling-Avg} layer. For the normal configurations, we can also observe sufficient diversity between different benchmarks, where the inter-layer diversity can be observed. Taking the normal configuration Cfg.\,A as an example, the average reuse distance of \emph{Conv.} (i.e., 3468) is two orders of magnitude larger than that of \emph{FC} (i.e., 24)\zadd{, see Figure~\ref{fig:avg_reuse_distance}}.

\textbf{Memory characteristics.} Figure~\ref{fig:avg_reuse_distance} shows the average reuse distances of all microbenchmarks, where the reuse distance ranges from 1.2 to 3.8E+7. For the normal configurations of most benchmarks, the reuse distances are less than 10 (e.g., \emph{Pooling}, \emph{ReLU}, and \emph{LRN}), and thus designing an architecture with on-chip local memory (i.e., cache or scratchpad memory) of $10 \times element\_size$ can significantly reduce off-chip memory accesses (except for the \emph{Conv.}, \emph{FC}, and \emph{Deconv}). In an extreme case such as Cfg.\,F of \emph{Conv}, the reuse distance is more than $10^6$, and an on-chip memory buffer would be more than 10MB for holding all the data, which is prohibitively costly with existing SRAM technology. In this case, the on-chip eDRAM employed in DaDianNao~\cite{Chen2015} and on-die 3D-stacked DRAM~\cite{Pawlowski11Hotchips} would be a potential solution for alleviating the penalty of vast off-chip memory accesses caused by large reuse distances.

\begin{figure}[!t]
   \vspace{-0.3cm}
  \centering
    \hspace*{-0.6cm}
  \includegraphics[width=1.1\columnwidth]{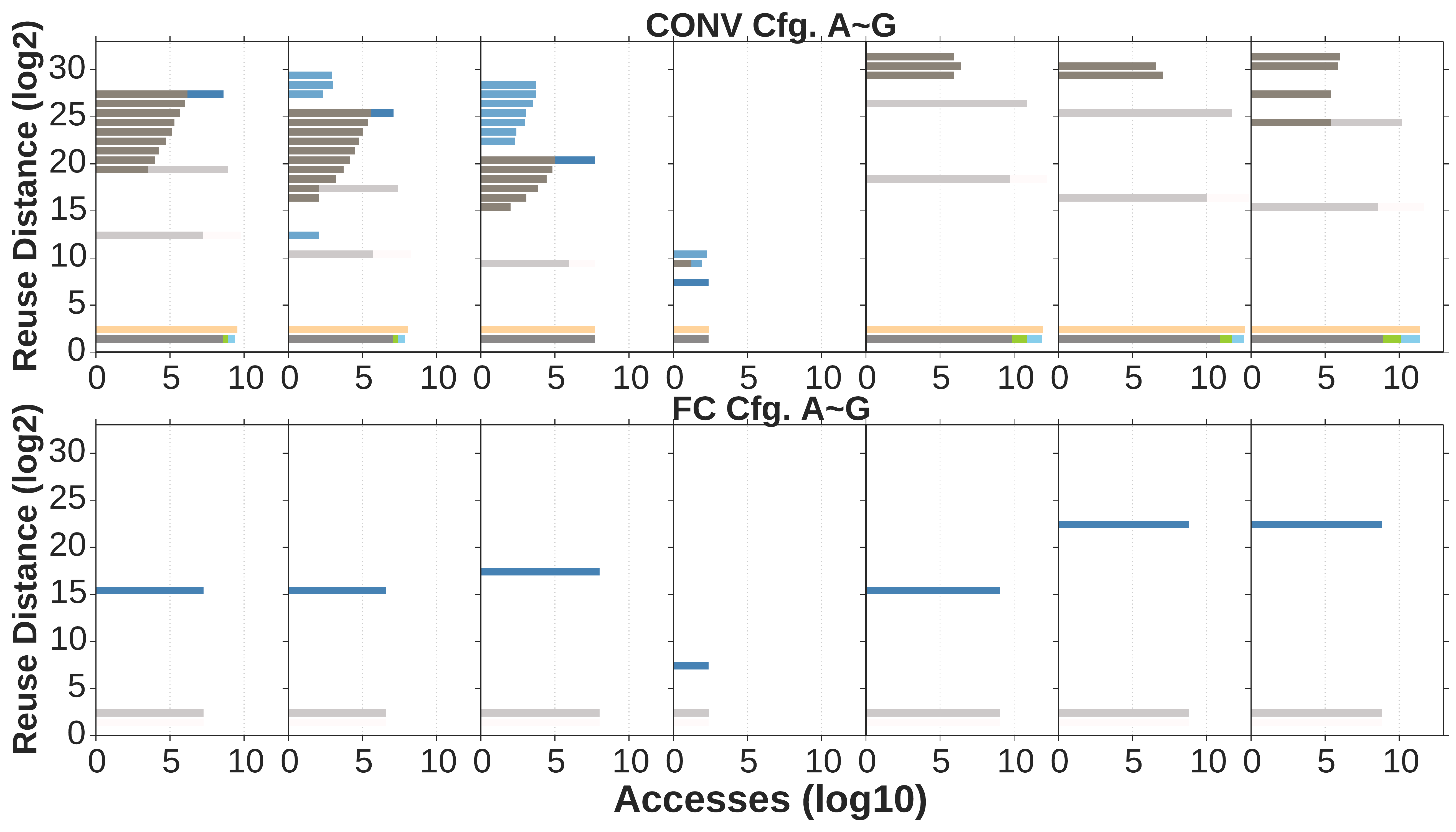}
  \vspace{-0.8cm}
  \caption{Reuse distances of all configurations of~\emph{Conv.} and~\emph{FC}.}
  \label{fig:reuse_distance}
\vspace{-0.6cm}
\end{figure}

Figure~\ref{fig:reuse_distance} further shows the reuse distances of all configurations of \emph{Conv.} and \emph{FC}, which have a relatively large average reuse distance. For the \emph{Conv.} layer, the reuse distance ranges from $2^1$ to $2^{29}$ for the normal configurations. While for the extreme configurations, the reuse distance would exceed $2^{30}$, posing a higher challenge on the underlying memory hierarchy. For the \emph{FC} layer, the reuse distances can be roughly classified into three groups: $2^2$, $2^{17}$, and $2^{22}$. Thus, for the \emph{FC} layer, it is intuitive to customize the memory hierarchy with three levels to accommodate the reuse distances listed above.

\begin{figure}[!t]
  \centering
    \hspace*{-0.6cm}
  \includegraphics[width=1.1\columnwidth]{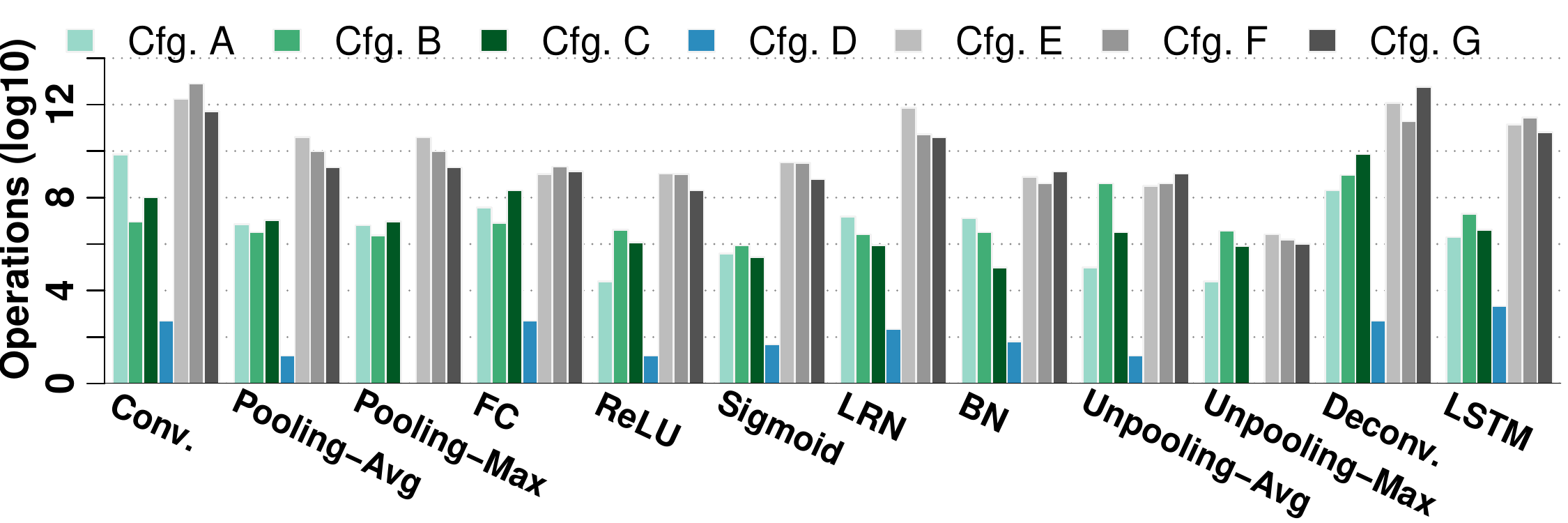}
  \vspace{-0.8cm}
  \caption{The number of operations of microbenchmarks.}
  \label{fig:ops}
\vspace{-0.6cm}
\end{figure}

\textbf{Computation characteristics.} Figure~\ref{fig:ops} shows the number of operations (Ops) of all benchmarks, where the maximal number of operations could be 8E+12 (Cfg.\,F of \emph{Conv.}). For the normal configurations, most Ops are less than $10^{8}$, while the Ops reach about $10^{10}$ in some cases. Given a specific benchmark, the Ops also varies significantly to well demonstrate the diversity. A more detailed analysis is the ratio of operations to memory access (i.e., OpMem). The layer with the highest value is the Cfg.\,F of the \emph{Conv.} layer ($>33$), while the Cfg.\,F of the \emph{FC} layer is only 0.039. This observation is in accordance with the intuition that the \emph{Conv.} layer is compute intensive and the \emph{FC} layer is relatively memory intensive.

\begin{figure}[!t]
    \vspace{-0.5cm}
  \centering
  \hspace*{-0.6cm}
    \includegraphics[width=1.1\columnwidth]{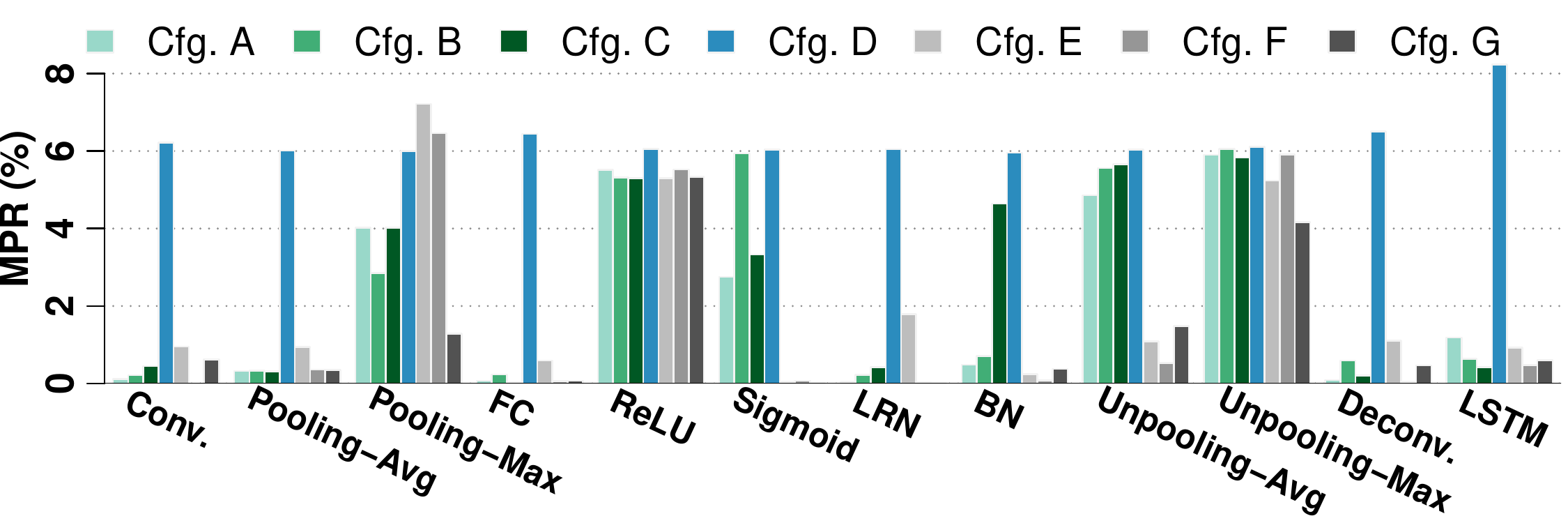}
  \vspace{-0.8cm}
  \caption{MPR of microbenchmarks.}
  \label{fig:mpr}
  \vspace{-0.3cm}
\end{figure}

\textbf{Control characteristics.} Figure~\ref{fig:mpr} illustrates the MPR of all benchmarks. An interesting observation is that the smallest configuration always has the highest MPR compared with other configurations, as the number of computations is much less than that of others. Another observation is that for the \emph{Pooling} layer, the MPR values of the average pooling and max pooling are completely different. For example, for the normal configuration Cfg.\,A, the MPR of average pooling is less than 0.5\%, while the MPR of max pooling is about 4\%. The underlying reason is that max pooling consists of comparison operations that do not exist in average pooling.

\subsubsection{Architecture perspective}
\vspace{-0.1cm}

From the architecture perspective, the benchmarks should be able to explore the boundary of the processing ability of intelligence processors. This is achieved by using the extreme configurations (Cfg.\,E$\sim$G) of each benchmark. In addition, the benchmarks should be timely and reflect future trends in the intelligence processors.

\textbf{Stress testing.} According to Figure~\ref{fig:avg_reuse_distance}, the average reuse distances of extreme configurations are much larger than those of the normal configurations, exhibiting different behaviors on conventional architectures. Given the \emph{Conv.} layer as an example, on a specific CPU (i.e., Intel i5-3470), the L2 cache miss rate of Cfg.\,F (67\%) is 8.4x larger than that of the normal configuration as Cfg.\,C (8\%). Moreover, the average reuse distances of extreme configurations in \emph{Conv.}, \emph{Deconv.}, and \emph{LSTM} are larger than those of others, which puts extreme pressure on the memory hierarchy for these benchmarks. For example, the L3 cache miss rate of Cfg.\,C in \emph{Conv.} is 1.26x larger than that of Cfg.\,C in \emph{ReLU}.

In Figure~\ref{fig:ops}, the number of operations of the most extreme configurations is larger than that of the normal configurations, except for \emph{Unpooling-Max}. The reason is that an extremely large unpooling window size leads to few operations. Regarding the ratio of operations to memory access (OpMem), the extreme configurations are quite different from the normal configurations. Taking the \emph{Conv.} layer as an example, the OpMem of Cfg.\,F is 33.1, while the OpMem of Cfg.\,A is only 0.36. In this case, the memory system is more crucial for extreme configurations than the normal ones.

Regarding the control characteristics in Figure~\ref{fig:mpr}, the extreme configurations do not exhibit significantly different behaviors. For the \emph{ReLU} layer, the MPR of extreme configurations is almost the same as that of the normal configurations. A potential reason is that the number of comparison operations increases in proportion to the input size, leading to a comparable branch and misprediction ratio regardless of the input size.

\textbf{Future trends.}
We expect that there are at least two trends for deep learning. The first is that more algorithms and network structures will be proposed for achieving \emph{higher} accuracy on solving \emph{more} generalized problems. As our benchmarks already have sufficient diversity in terms of computation, memory, and control behaviors, the problem can be alleviated to a certain extent. Besides, we will also update the \textsc{BenchIP} benchmark suite frequently to keep pace with the rapid advance of deep learning by either adding new benchmarks or eliminating out-of-date benchmarks. The second trend is that the network model tends to be more compact by sacrificing accuracy for performance/energy efficiency. The most notable techniques include the \emph{sparse} model and \emph{low-precision} models (e.g., models with an 8-bit fixpoint and even 1-bit values). To reflect this trend, for the microbenchmarks, we provide sparse and 16-bit models for \emph{Conv.} and \emph{FC} that requires weight values. For the macrobenchmarks, we provide sparse models for LeNet-5, AlexNet, and VGG, as they are the basis of many others networks (e.g., Faster-RCNN and S2VT). The research community is very active in considering both accuracy and architectural improvement, including pruning and compression. In the future, we will update the benchmarks by providing more sparse networks and models.

\vspace{-0.2cm}
\section{Benchmarking Methodology}
\label{sec:methodology}
In this section, we first present the guidelines and overview of the proposed benchmarking methodology in \textsc{BenchIP}. Then, we detail the benchmarking framework.

\begin{figure}[!t]
  \centering
  \vspace{-0.5cm}
  \includegraphics[width=0.75\columnwidth,height=3cm]{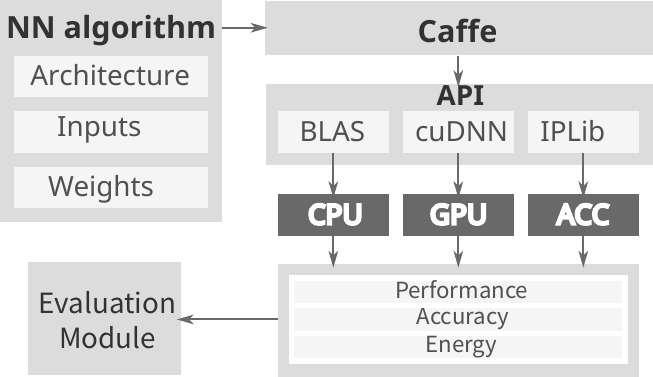}
  \vspace{-0.3cm}
  \caption{The overall framework of \textsc{BenchIP}'s benchmarking methodology.}
  \label{fig:framework}
    \vspace{-0.6cm}
\end{figure}

\subsection{Guidelines and overview}

\textsc{BenchIP} has two main objectives. The first is to provide a reference toolset for comparing emerging intelligence processors in a fair and easy-to-use fashion. The second is to facilitate identifying performance/energy bottlenecks for potential optimization. The benchmarking methodology should follow at least the following guidelines:

\begin{itemize}[leftmargin=0.5cm,topsep=0cm,itemsep=0.0cm]
\item \textbf{Creditability.} The benchmarking methodology should replicate the real application scenario.
\item \textbf{Portability.} The benchmarking framework should be portable across different platforms, i.e., CPUs, GPUs, and accelerators.
\item \textbf{Fairness.} The benchmarking methodology should provide fair comparison through specifying the rules and metrics.
\end{itemize}

To meet the above requirements, we offer a general software stack with a portable interface for different intelligence processors, and the software stack is a replicate of the most widely used industrial software stack, i.e., the stack consisting of high-level programming model, high-performance library, and device driver. The fairness is mainly guaranteed by specifying several benchmarking rules and specially designed evaluation metrics.

  \vspace{-0.2cm}
\subsection{Benchmarking framework}
In Figure~\ref{fig:framework}, the benchmarking framework is built upon Caffe~\cite{Jia14Caffe}, where all necessary layers and networks are implemented. \zadd{Currently, we support Caffe firstly because it is a widely used production-quality framework. We will support other software frameworks progressively.} The descriptions of network architecture, input datasets, and pre-trained synaptic weights are treated as standard inputs. The Caffe framework interfaces with the underlying high-performance libraries for different architectures---for example, BLAS for CPUs and cuDNN for GPUs. Given a new hardware accelerator, the corresponding library should be offered, so as to be integrated into the Caffe framework. The platform should be able to report the performance and energy. Also, the entire framework should report the accuracy of the neural networks. Finally, these individual results are sent to the evaluation module for producing final evaluation results.

\rmk{RD: It is not clear from the paper that how much of the results are actually caused by the Caffe framework itself (as opposed to the actual workload), and how much of the findings still hold if the workload runs on a different NN learning framework such as Torch or TensorFlow.}

\textbf{Benchmark specification.} In \textsc{BenchIP}, the following files are provided. (1) Configuration file. Since the benchmarking framework is built upon Caffe, the same configure file is used for describing the network architecture. In more detail, we provide the Caffe-compatible \textit{.prototxt} file to characterize both single-layer and entire networks, specifying parameters of each layer and connections between layers. Note that single layers are also considered as independent networks composed of a data layer and the tested layer. (2) Learned parameters (e.g., weight and bias for Conv and FC, $\gamma$ for BN). For each single layer and entire network, the \textit{.caffemodel} file, where the parameters of every weight layer in that network are stored, is offered. (3) Reference output. \textsc{BenchIP} offers reference output to measure the precision of tested intelligence processors. Regarding microbenchmarks, we provide input data and reference output data, and the difference between the reference output and the computed output is measured using Mean Squared Error (MSE). Regarding macrobenchmarks, since we select networks from different application scenarios, the evaluated metrics vary as well. Intuitively, their original metrics are employed. For example, we employ the \emph{Top-5} error rate for measuring the networks that aim at image classification (e.g., AlexNet) and METEOR\cite{Denkowski2014} for networks targeting video captioning (e.g., S2VT).

\textbf{Library interface.} We provide a standard library interface, allowing convenient evaluation of different architectures with the Caffe-style framework. It comprises a set of high-level function interfaces implemented in C/C++, each of which corresponds to a basic operation in neural networks. Given a new intelligence processor, the standard interface can be overriden with its own library. Listing \ref{list:apiexample} presents examples of such interfaces and how to implement them with user-provided library. The \texttt{accConvolutionLayer} and virtual function \texttt{Forward\_acc()} are provided by our benchmarking framework. The virtual function is then implemented by the user-provided library function \texttt{accConvolutionForward()}. We also provide an interface for fusing multiple layers---\texttt{accFusionLayer::Forward\_acc()}---since some processors could be able to obtain better performance by eliminating boundaries between consecutive layers. As shown in Listing~\ref{list:apiexample}, the standard interface is implemented with the user-provided \texttt{accMultiLayerForward()} so as to optimize the combined execution of the convolutional and pooling layers.

\zadd{\textbf{Evaluation metrics.} There are four important design considerations for intelligence processors: performance, energy, area, and accuracy. These are deployed as the evaluation metrics. Performance and energy should be reported when running the benchmarks. On CPUs and GPUs, such metrics can be collected using PAPI~\cite{Mucci99PAPI} and nvsim, respectively. On the accelerator, if it is a simulation prototype, the corresponding simulator should be able to report performance, energy, and area. The collection of the accuracy metric is built into the framework as well.}

\begin{figure}[t]
  \vspace{-0.7cm}
\noindent\begin{minipage}[!t]{\columnwidth}
\vspace{-0.2cm}
\begin{lstlisting}[language=c++,caption={Example of Caffe implementation},belowcaptionskip=0.5cm,label={list:apiexample},emph={accConvolutionLayer,accFusionLayer},emphstyle=\color{dkgreen}]
// standard interface for a single layer
void accConvolutionLayer::Forward_acc(...) {
  ...
  accConvolutionForward(...); // user-provided lib
  ...}
// standard interface for fused layers
void accFusionLayer::Forward_acc(...) {
  ...
  accMultiLayerForward(Conv, Pool, ...); // user-provided lib
  ...}
\end{lstlisting}
\vspace{-0.8cm}
\end{minipage}
\vspace{-0.2cm}
\end{figure}

\zdel{\rmkdone{ZD: whether should the synthesized score which synthesized all efficiency scores should be introduced?}}
\zadd{For microbenchmarks, such metrics already cover various design tradeoffs, and they are sufficient for optimization. For macrobenchmarks, we also provide several synthesized scores for better evaluating design tradeoffs balancing, from the \emph{efficiency} perspective. Such efficiency includes \emph{operations per Joule} as energy efficiency, \emph{operations per second} as computation efficiency and \emph{accuracy affected by area savings} as silicon efficiency.}

\zadd{With such direct and synthesized metrics, it is relatively easy to conduct platform comparisons and optimizations in multiple dimensions, e.g., performance, energy, and various efficiencies. This allows architects to consider potential designs for flexibly balancing different tradeoffs. Given two platforms, macrobenchmarks can be deployed on both platforms to obtain the overall evaluation results. With there results, end-users can perform an overall comparison of these two platforms within their interests. For the weak platform, the architects can leverage microbenchmarks to conduct a problem diagnosis. For the better one, microbenchmarks can be used for further improvements.}

  \vspace{-0.3cm}
\section{Benchmarking with BenchIP}
\label{sec:experiments}

\rmk{Dispense the section into places where readers are interested.}

In this section, we show how to benchmark IPs by using \textsc{BenchIP}. We first measure eight IPs (three CPUs, three GPUs and two accelerators) with macrobenchmarks to obtain the overall scores. Then, we delve into the details of the execution efficiency of IPs using the fine-grained microbenchmarks.

\subsection{Evaluated IPs}

\textbf{CPU.} We select three types of CPUs---CPU-E, CPU-D and CPU-S---for different scenarios. CPU-E is an embedded processor, i.e.,  4xA57 Atlas/2MB L2. CPU-D is a desktop processor, i.e., Intel i5-3470@3.2GHz with 16GB memory. CPU-S is a server processor, i.e., Intel E5-2620 v2@2.1GHz with 64GB memory.

\textbf{GPU.} Similarly, we select three GPUs with different levels of capabilities: GPU-E, GPU-D, and GPU-S. GPU-E is a embedded GPU, i.e., Nvidia Maxwell GeForce~\cite{maxwell} (256 CUDA cores, 512 GFlops with FP32). GPU-D is a desktop GPU, i.e., GeForce GTX950~\cite{950} (640 CUDA cores, 2GB GDDR5, 1.57 TFlops). GPU-S is a server GPU, i.e., Nvidia Tesla K40~\cite{k40} (2880 processors, 12GB GDDR5, 4.29 TFlops).

\textbf{ACC.} We select two prototypes of ASIC IPs---ACC-1 and ACC-2---by carefully reimplementing DianNao~\cite{Chen14ASPLOS} and Cambricon~\cite{Liu2016a}, respectively, with exactly the same parameters in the papers. The reason for selecting DianNao and Cambricon as prototypes of ACC-1 and ACC-2 is twofold. First, DianNao is a notable accelerator that has moderate performance and cost that supports basic matrix/vector operations in neural networks. Second, Cambricon is one of the few accelerators that supports various kinds of neural network algorithms but still has performance comparable with the state-of-the-art high-performance accelerator, DaDianNao~\cite{Chen2015}. Thus, ACC-1 and ACC-2 can cover various scenarios.

\begin{figure}[!t]
    \vspace{-0.2cm}
  \centering
  \includegraphics[width=1.05\columnwidth]{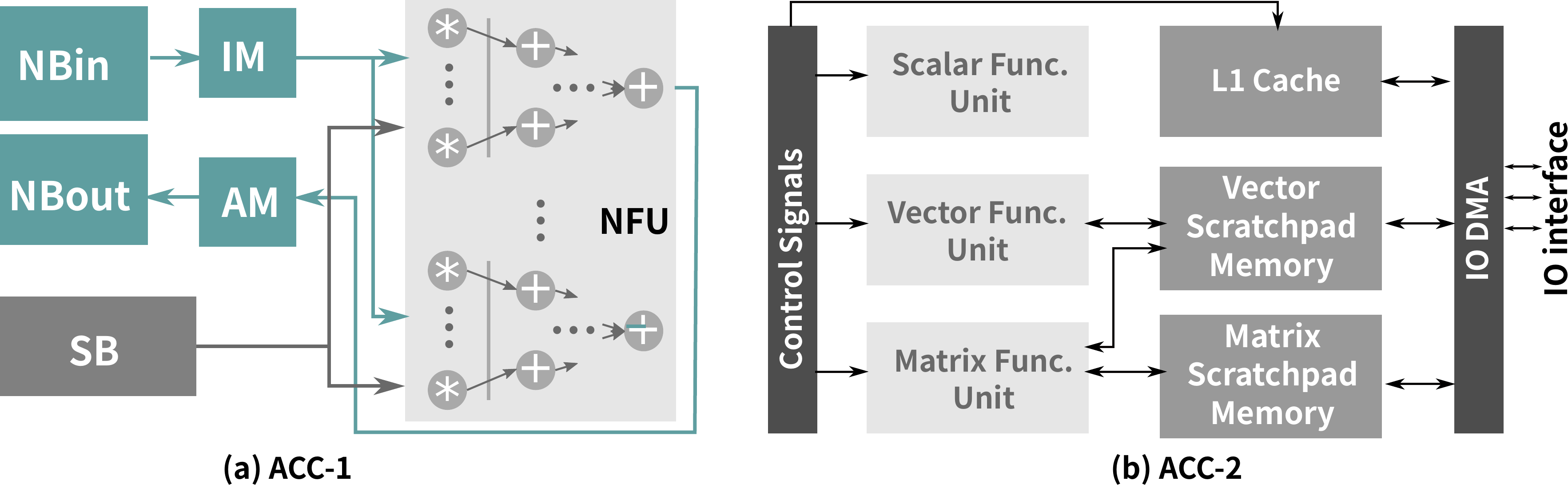}
  \vspace{-0.7cm}
  \caption{The frameworks of ACC-1 and ACC-2.}
  \label{fig:acc}
  \vspace{-0.5cm}
\end{figure}

In Figure~\ref{fig:acc}, we present the framework of both ACC-1 and ACC-2. ACC-1 consists of a neural functional unit (NFU), an input/output neuron buffer (NBin/NBout), a synaptic buffer (SB), an indexing module (IM), and an assemble module (AM). Note that we modify the architecture of DianNao to produce ACC-1, by adding several features such as sparsity (i.e., IM). ACC-2 consists of a matrix function unit (MFU), a vector function unit (VFU), a scalar function unit (SFU), an address generation unit (AGU), and companied scratchpad memory.

\vspace{-0.2cm}
\subsection{Benchmarking with macrobenchmarks}

\begin{figure}[t]
    \vspace{-0.2cm}
    \centering
      \hspace*{-0.6cm}
  \includegraphics[width=1.1\columnwidth]{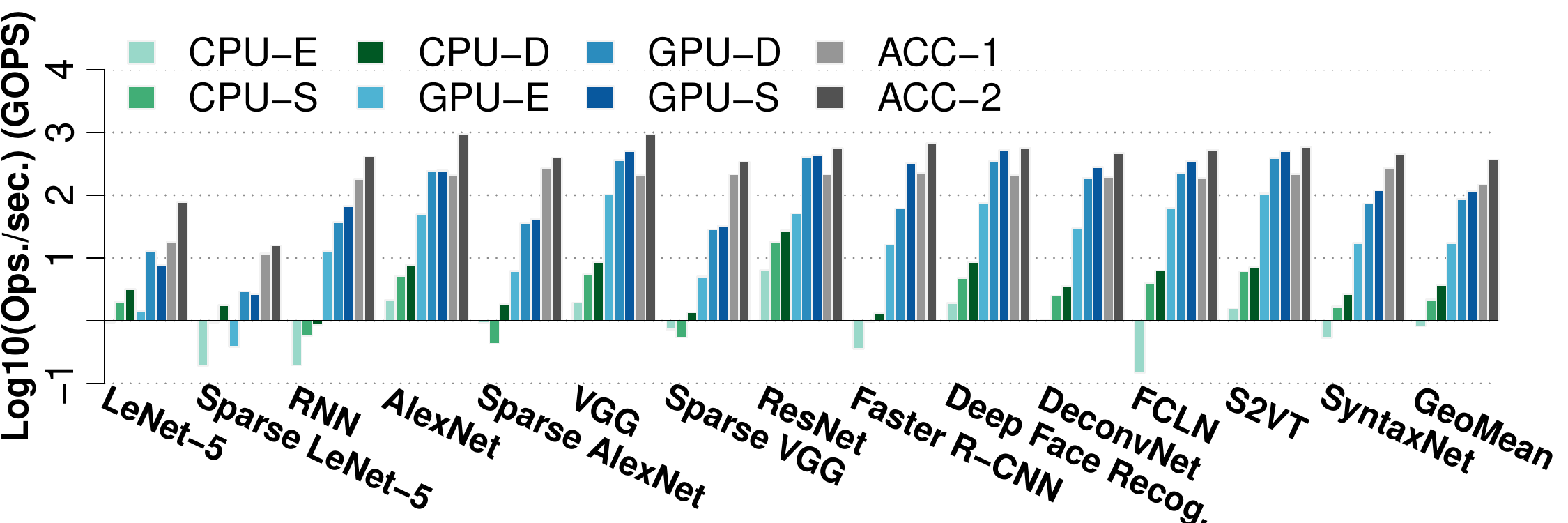}
  \vspace{-0.7cm}
  \caption{Giga operations per second.}
  \label{fig:gflops}
  \vspace{-0.2cm}
\end{figure}

\begin{figure}[t]
  \vspace{-0.2cm}
  \centering
  \hspace*{-0.6cm}
  \includegraphics[width=1.1\columnwidth]{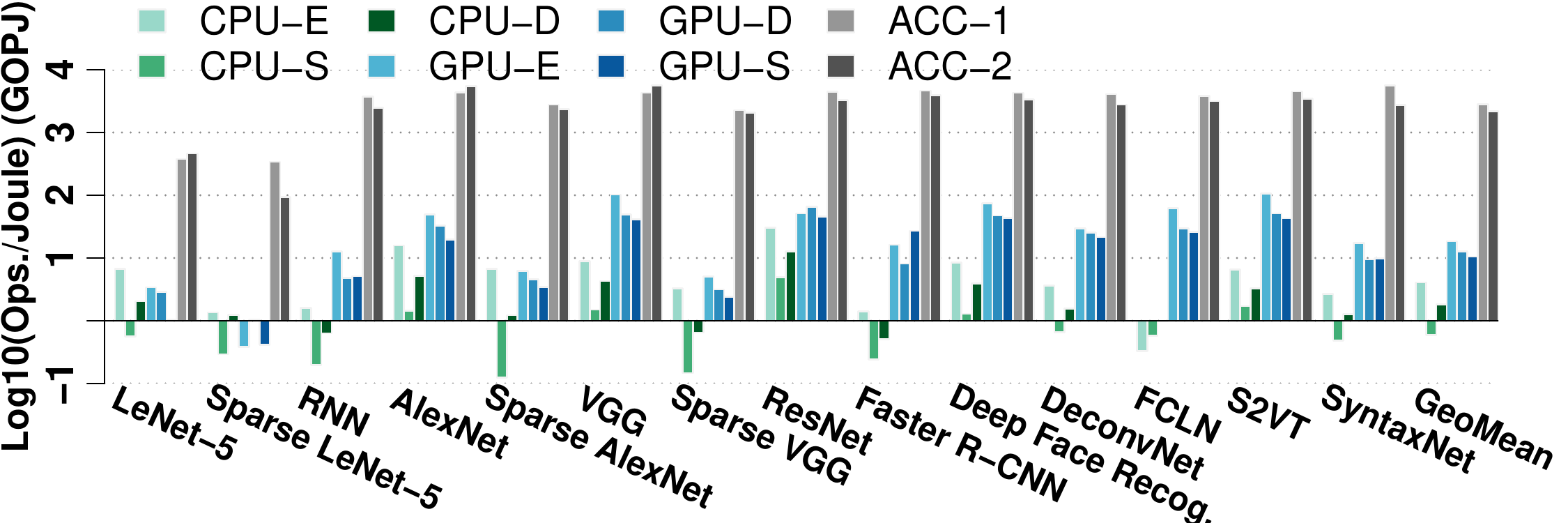}
  \vspace{-0.7cm}
  \caption{Giga operation per joule.}
  \label{fig:gflopj}
  \vspace{-0.2cm}
\end{figure}

\begin{figure}[t]
  \centering
  \hspace*{-0.2cm}
  \includegraphics[width=1.1\columnwidth]{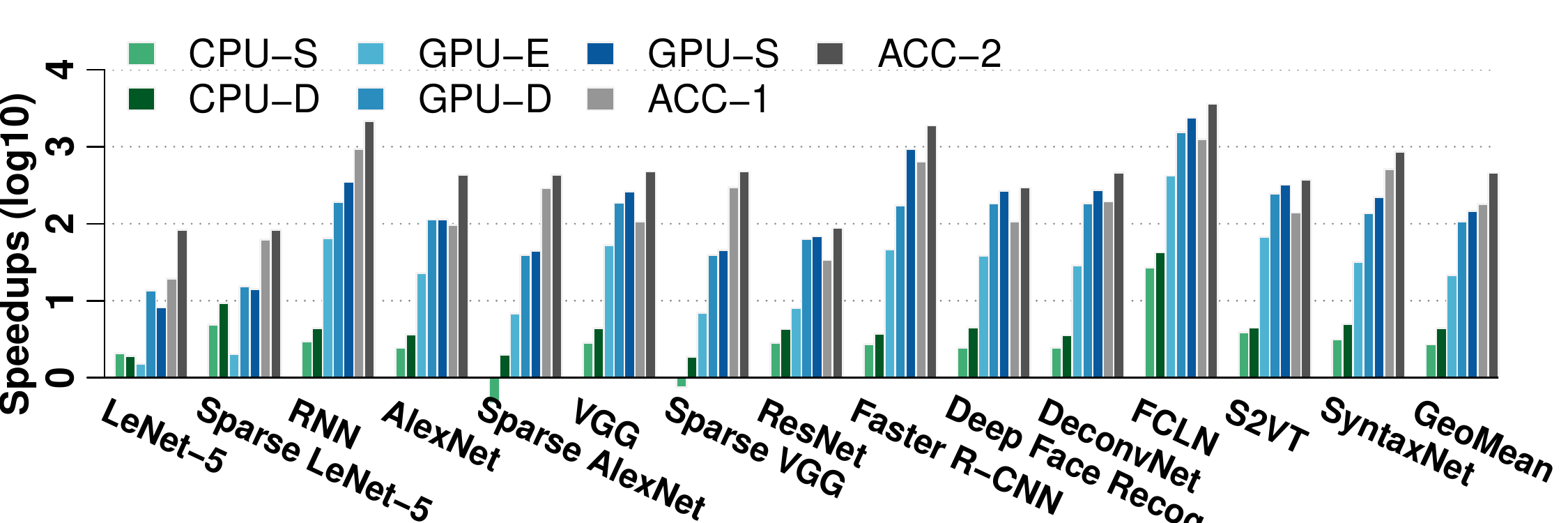}
  \vspace{-0.8cm}
  \caption{Speedups of evaluated IPs with macrobenchmarks (Normalized to CPU-E).}
  \label{fig:speedup}
  \vspace{-0.2cm}
\end{figure}

\begin{figure}[t]
      \vspace{-0.6cm}
  \centering
  \hspace*{-0.6cm}
  \includegraphics[width=1.1\columnwidth]{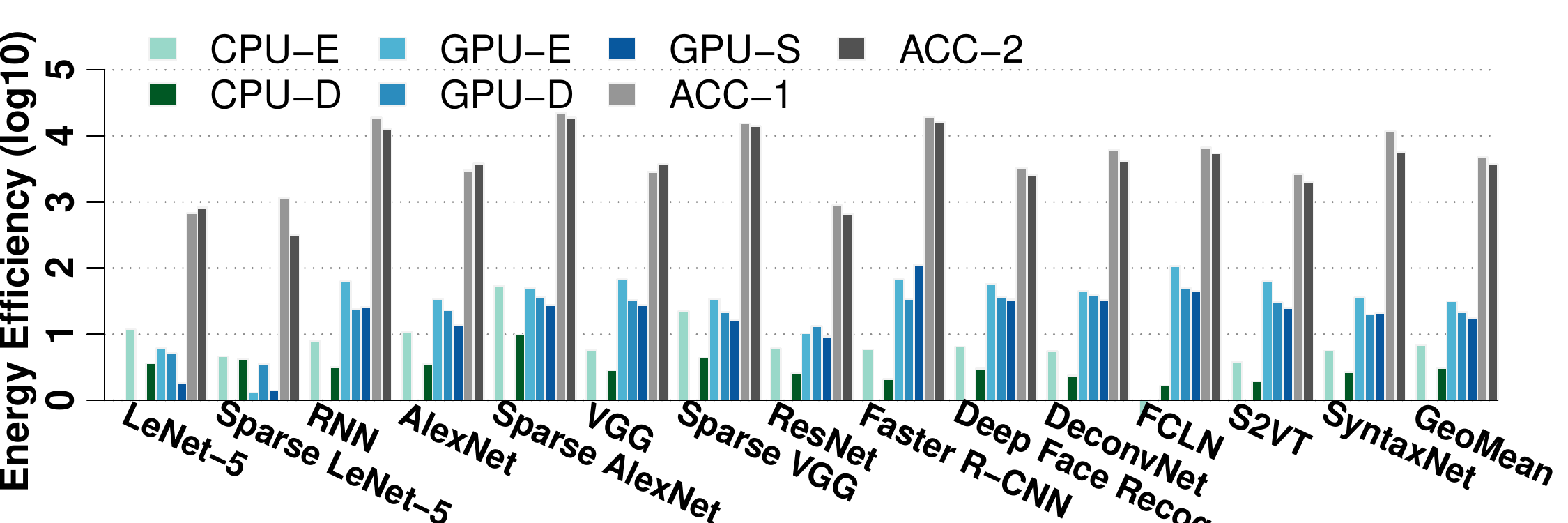}
    \vspace{-0.8cm}
  \caption{Normalized energy consumption of evaluated IPs with macrobenchmarks (Normalized to CPU-S).}
  \label{fig:energy}
    \vspace{-0.5cm}
\end{figure}

\zadd{In Figure~\ref{fig:gflops} and~\ref{fig:gflopj}, we report the synthesized scores from the efficiency perspective, i.e., energy efficiency using \emph{giga operations per Joule} (GOPJ) and performance efficiency using \emph{giga operations per second} (GOPS). Further, we report the fine-grained metrics, i.e.,~\emph{performance} and~\emph{energy}, in Figure~\ref{fig:speedup} and~\ref{fig:energy}, respectively. We focus on performance and energy because area and accuracy basically do not vary across different macrobenchmarks and IPs, respectively. Thus, we have the following observations. Here also note that the overhead of the Caffe software stack is small, less than 2\% of the total runtime on CPUs/GPUs. And for all the customized intelligence processors, we fetch end-to-end measured data for benchmarking. Thus, the results would not be affected by software stack.}

\zadd{\textbf{Observation \#1.}~\emph{Embedded CPUs trade performance for silicon cost reduction better than energy efficiency.} Lightweight embedded CPUs are well known for being suitable for power limited scenarios such as embedded systems, mainly because of their low energy costs. Interestingly, as we observed, an embedded CPU, i.e., CPU-E, improves silicon area utilization better than energy efficiency. CPU-E achieves 9.22x and 11.11x better silicon area utilization but only 6.91x and 2.23x better efficiency in utilizing every joule for computation (GOPJ) than CPU-S and CPU-D, 5.03x and 3.12x slower in performance, respectively. \zdel{\emph{Therefore, embedded CPUs, instead of desktop and server CPUs, seem to be the best option for intelligence processing in limited scenarios or less intensive networks.}}}

\zadd{\textbf{Observation \#2.}~\emph{GPUs still improve energy efficiency as well as provide high performance efficiency.} GPUs let applications exploit parallelism with tremendous computation units and memory bandwidth, which motivated their widely usage in intelligence processing. \note{As observed, we confirmed that GPUs still leverage the energy for more operations~\cite{Keckler2011}}. On average, GPU-E, GPU-D and GPU-S achieve \note{4.65x}, \note{23.14x}, and \note{31.55x} better computation efficiency than CPU-D while still achieving certain improvements in energy efficiency, i.e., \note{5.74x}, \note{7.02x}, and \note{10.15x}, respectively. Interestingly, GPU-D has the best utilization of such a tradeoff. Regarding the efficiencies, GPU-D achieves relatively high performance and energy efficiency, i.e., 69.17\% of GPU-E (122.42\% of GPU-S) in GOPJ and 73.33\% of GPU-S (497.99\% of GPU-E) in GOPS. When detailed in energy costs and execution times, compared to GPU-S, GPU-D has 0.84x energy consumption and 1.41x slower performance with 40.64\% area cost. Moreover, GPU-D achieves 4.91x speedup over GPU-E and 1.47x energy costs, with 1.43x more area cost.\zdel{~\emph{Therefore, desktop GPUs, instead of embedded and server GPUs, seem to be the best option for intelligence processing as their better efficiency in performance improvement.}}}

\zadd{\textbf{Observation \#3.}~\emph{Customized accelerators provide significantly better energy efficiency with a relatively lower performance efficiency improvement and silicon area cost reduction.} We observed that ACC-1 and ACC-2 achieve \note{1.69x} and \note{4.26x} performance efficiency but \note{220.10x} and \note{170.80x} energy efficiency when compared against GPU-D, and \note{39.21x} and \note{98.55x} performance efficiency but \note{1545.78x} and \note{1199.56x} energy efficiency when compared against CPU-S. Regarding both performance and energy efficiency, ACC-1 and ACC-2 have \note{higher utilization} in silicon area, i.e., \note{44.31\%/4.00\%} and \note{390.06\%/35.15} of smallest/largest CPU (CPU-E/CPU-S), \note{18.77\%/165.42\%} and \note{1.14\%/10.03\%} of smallest/largest GPU (GPU-E/GPU-S).\zdel{~\emph{Therefore, customized accelerators seem to be the best platform for almost all scenarios, from embedded system to server end, with tremendous efficiency improvements in energy, performance and silicon area utilization.}}}

\zadd{\textbf{Observation \#4.}~\emph{ACC-1 is more efficient than ACC-2, which complies with the intuition that with sparse support the accelerator benefits from such sparsity}. Interestingly, ACC-1 is more efficient than ACC-2 in GOPJ (1.29x) but 39.79\% inefficient in GOPS. Moreover, due to the sparsity, ACC-1 is slightly more efficient in GOPJ, where ACC-1 has a 2.99x speedup on average with sparse support and ACC-2 gains nothing as it processes no differently on sparse and dense networks. Furthermore, note that despite the very small network LeNet-5, both CPUs and GPUs cannot leverage the benefit of computations and data amount reduction due to sparsity, i.e., 2.09x and 4.15x slower when compared against dense version. When detailed in energy costs and execution times, ACC-1 performs 2.51x slower than ACC-2 but costs 1.29x less in energy and has far less area cost, averaging on all macrobenchmarks.}

\zadd{In short, while using~\textsc{BenchIP}, we made the following observations from the efficiency perspective:(1) embedded CPUs trades performance efficiently for small area better than low power, (2) GPUs focus on performance improvement but still improve the energy efficiency, and desktop GPUs have the best tradeoff, (3)customized accelerators are suitable for almost all scenarios, and (4) accelerators with sparse support are much better than their dense counterparts.}

\zdel{In short, by using~\textsc{BenchIP}, we made the following observations from the efficiency perspective: (1) embedded CPUs seem to be the better option than desktop and server CPUs, (2) desktop GPUs are more efficient than embedded and server GPUs, and (3) accelerators with sparse support are much better than dense counterparts.}

    \vspace{-0.2cm}
\subsection{Benchmarking with microbenchmarks}

\begin{figure}[t]
      \vspace{-0.6cm}
  \centering
  \hspace*{-0.6cm}
  \includegraphics[width=1.1\columnwidth]{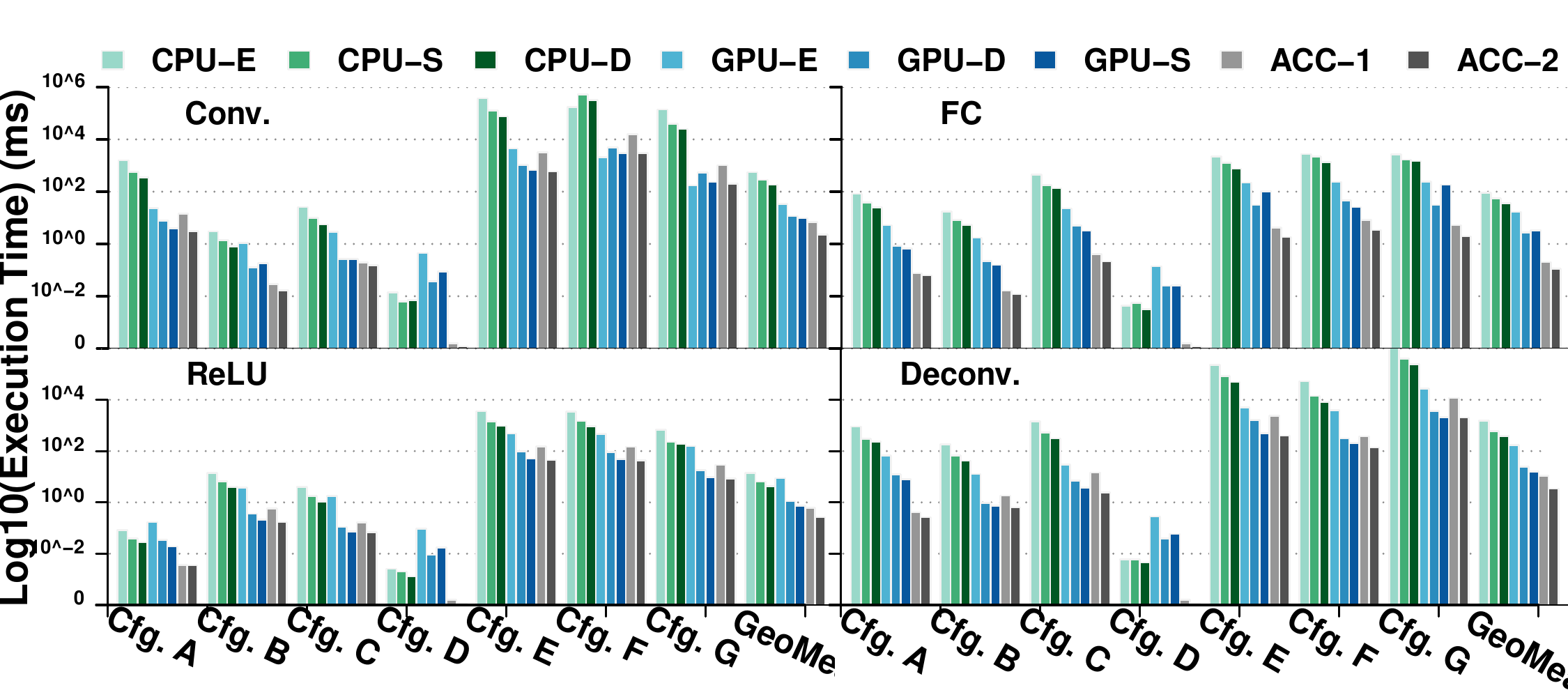}
    \vspace{-0.9cm}
  \caption{Performance of IPs on microbenchmarks (~\emph{Conv., FC, ReLU} and~\emph{Deconv}).}
  \label{fig:micro_all}
  \vspace{-0.5cm}
\end{figure}

For the platforms with relatively low/high scores, we can study the reasons for the inefficiency/efficiency with microbenchmarks. As illustrative examples, we further evaluate especially CPU-D, GPU-E, and ACC-2 with microbenchmarks, as they perform worse on macrobenchmarks. In Figure~\ref{fig:micro_all}, we shown the performance of IPs on certain microbenchmarks, i.e.,~\emph{Conv., FC, ReLU}, and~\emph{Deconv.}, which cover all three computation patterns (corresponding to RD, RD, EW, and EL).

Regarding observation \#1, the desktop-level CPU-D achieves better performance than both CPU-E and the server end CPU-S, i.e., 5.03x and 1.61x on average, respectively, as shown in Figure~\ref{fig:speedup}. Averaging on microbenchmarks with all the configurations, CPU-D achieves results consistent with those on macrobenchmarks, with 3.41x and 1.48x speedups over CPU-E and CPU-S. Also CPUs perform uniformly on normal or larger configurations (Cfg.\,A$\sim$Cfg.\,C, Cfg.\,D$\sim$Cfg.\,G), where the speedups vary from 3.36x/1.50x to 4.07x/1.61x.

Regarding observation \#2, the embedded system GPU, GPU-E, does not always have a shorter execution time than the CPUs, especially on small networks (~\emph{LeNet-5} and~\emph{sparse LeNet-5}). With the microbenchmarks, we observe that GPU-E has 0.02x speedup on average compared against CPU-D on Cfg.\,D which is the extremely small cases. GPUs cannot take full advantage of their high computational power because the small computational kernels map poorly on their hundreds and thousands of threads.

Regarding observation \zdel{\#3}\zadd{\#4}, ACC-2 achieves the best performance on every macrobenchmark and it is the same for all the microbenchmarks, i.e., a 2.2x speedup averagely. Moreover, compared against ACC-1, ACC-2 performs better on large configurations (3.19x speedup) than normal configurations (1.80x speedup) where its high parallelism and throughput (peak performance of 2114 GOPS) can be well leveraged.

Additionally, ACC-1 achieves better performance than all the CPUs and GPUs on most CNNs and DNNs but fails on complex and irregular networks such as~\emph{ResNet} and~\emph{DeconvNet}. It is because ACC-1 is designed to efficiently process CNNs/DNNs. Thus, ACC-1 obtains a performance benefit over GPU-S on layers commonly used in CNNs and DNNs, i.e., \emph{Conv.} (1.43x) and \emph{FC} (16.25x).

ACC-1 can be very effective on small networks, as it performs 217.78x faster than GPU-S on Cfg.\,D. On normal configurations such as  Cfg.\,A$\sim$C, ACC-1 can still keep its efficiency with a 1.71x speedup. However, on heavy workloads, GPUs can well leverage their high parallelism. Thus, ACC-1 cannot exceed GPU-S (0.65x speedup on Cfg.\,D$\sim$\,G).

\vspace{-0.3cm}
\zadd{\section{Discussion}}
\zadd{\textbf{Overall Score.} In current version, we do not provide an overall score for different IPs. The reason is two-fold. First, most architecture designs involve a multi-dimension optimization process where architects make decisions on a series of tradeoffs that leads to a final design for some specific targets. Thus, in such case detailed evaluation results can be used directly for design improvement and an overall score would be meaningless. Second, overall scores appropriate for various platforms, e.g., CPU, GPU, FPGA, DSP, and customized accelerators do not exist. One possible definition of an overall score could be the overall efficiency, i.e., synthesizing energy efficiency, performance efficiency, and silicon efficiency. For example, the overall efficiency equation can be defined as
\mbox{\(\footnotesize
  GeoMean({f(\frac{Ops}{Energy}) \times g(\frac{Ops}{Times}) \times h(\frac{Acc/Ref\_acc}{Area}}))\)}, where the $f(\cdot)$, $g(\cdot)$ and $h(\cdot)$ are potential scaling/mapping functions. However, due to the huge gaps among various platforms, those functions can only be user-defined or fitting equations learning from collected data after \textsc{BenchIP} is opened for public. Therefore, at present we only provide direct measured metrics and synthesized efficiency scores.}

\vspace{-0.3cm}
\section{Related Work}
We review related work from three aspects: neural network models, accelerators, and benchmarks.

\textbf{Models.} In recent years, deep learning has been applied to various scenarios and achieved great success. For image recognition, variants of CNNs, as powerful feature extractors, are applied to different tasks (e.g., image classification and face recognition)~\cite{Krizhevsky,Simonyan2015,He,Parkhi2015}. Image/video captioning adopts both CNN and LSTM for extracting features and generating sentences, respectively~\cite{Venugopalan15ICCV,Johnson2015}. In semantic segmentation, deconvolution and unpooling are used for reconstructing features~\cite{Noh2015}. In addition to image processing, deep learning also achieves great performance on speech recognition~\cite{graves2013speech}, and natural language processing~\cite{Andor2016}, where LSTM/RNN are the core learning models. Such diverse scenarios have led to a number of variants of neural network architectures and models that have seldom been considered in previous studies of neural network accelerators.

\textbf{Accelerators.} With the growing consensus that CPUs and GPUs, as the traditional platforms for running neural network models, are not able to provide high energy-efficiency for specific workloads like CNN, researchers began to seek the possibility of implementing neural networks on other platforms (e.g., FPGA~\cite{Zhang15FPGA} and ASIC~\cite{Chen14ASPLOS,Chen2015,Du2015,Shafiee2016,Chi2016,Han2016a}). The DianNao family~\cite{Chen14ASPLOS,Chen2015,Du2015} contains a series of neural network processors to support most operations of CNN and DNN.

Han~\emph{et al.}~\cite{Han2016a} designed EIE, which is a specially optimized architecture for sparse networks. Chi~\emph{et al.}\cite{Chi2016} designed PRIME, a PIM architecture which accelerates NN algorithms in ReRAM-based main memory. The above studies evaluate their processors on different sets of deep learning operations and configurations, making it impossible to compare two processors in a fair fashion.

\textbf{Benchmarks.} Though the exploration of intelligence processors continues to grow, a suitable benchmark for the evaluation and optimization of such hardware is still absent. BenchNN and DeepBench are two benchmarks that partially address this problem. However, due to their non-diversity and nonrepresentativeness, they cannot be used for designing and optimizing state-of-the-art intelligence processors. In the computer architecture community, several benchmarks have already been used. Chen~\emph{et al.}\cite{Chen2015} used 10 layers all extracted from AlexNet and two customized RBMs for evaluation. Chi~\emph{et al.}\cite{Chi2016} employed six networks, five extracted from LeNet-5 and one from VGG. Han~\emph{et al.} employed nine DNNs extracted from AlexNet, VGG, and Neural Talk\cite{Karpathy2015}. Shafiee~\emph{et al.}\cite{Shafiee2016} employed seven CNNs (four from VGG, three from MSRA\cite{He2015}), and two DNNs (one from DeepFace\cite{Taigman2014} and the other one from \cite{Le2012}). Due to the lack of a standard benchmark suite, employing such personalized benchmarks in the design of existing neural network accelerators is relatively casual, which is unsuitable for benchmarking intelligence processors.

\vspace{-0.35cm}
\section{Conclusions}
\label{sec:conclusion}
This paper proposes \textsc{BenchIP}, a benchmark suite and methodology for benchmarking intelligence processors. The benchmark suite of \textsc{BenchIP} consists of two sets of benchmarks, i.e., microbenchmarks and macrobenchmarks---for fair comparison and system optimization. The benchmark methodology is built upon an industrial software stack consisting of a high-level programming model, library, and device driver. We also provide \zdel{hierarchical}\zadd{evaluation} metrics for comprehensively reflecting the various characteristics of the intelligence processors being evaluated. \textsc{BenchIP} is employed to evaluate various hardware platforms, including CPUs, GPUs, and neural network accelerators. \textsc{BenchIP}\zadd{, which is already being used in several companies,} will be open-sourced soon to facilitate the design and evaluation of intelligence processors in a broad sense.

%\biography{photo.eps}%No empty line here
%{\bf Author Name} 
%Biography text here.

%\end{multicols}
%\end{CJK}

\begin{thebibliography}{99}

\bibitem{Krizhevsky}
A.~Krizhevsky, G.~E. Hinton, I.~Sutskever, and G.~E. Hinton, ``{ImageNet
  Classification with Deep Convolutional Neural Networks},'' {\em Advances In
  Neural Information Processing Systems}, pp.~1--9, 2012.

\bibitem{Simonyan2015}
K.~Simonyan and A.~Zisserman, ``{Very Deep Convolutional Networks for
  Large-Scale Image Recoginition},'' pp.~1--14, 2015.

\bibitem{He15CVPR}
K.~He, X.~Zhang, S.~Ren, and J.~Sun, ``Deep residual learning for image
  recognition,'' in {\em arXiv prepring arXiv:1506.01497}, 2015.

\bibitem{Venugopalan15ICCV}
S.~Venugopalan, M.~Rohrbach, J.~Donahue, R.~J. Mooney, T.~Darrell, and
  K.~Saenko, ``Sequence to sequence -- video to text,'' in {\em Proceedings of
  the 2015 International Conference on Computer Vision (ICCV'15)}, 2015.

\bibitem{Ossama14TASLP}
O.~Abdel-Hamid, A.~rahman Mohamed, H.~Jiang, L.~Deng, G.~Penn, and D.~Yu,
  ``Convolutional neural networks for speech recognition,'' {\em IEEE/ACM
  Transactions on Audio, Speech, and Language Processing}, vol.~22,
  pp.~1533--1545, 2014.

\bibitem{Akiko16ACL}
A.~Eriguchi, K.~Hashimoto, and Y.~Tsuruok, ``Tree-to-sequence attentional
  neural machine translation,'' in {\em Proceedings of the 54th Annual Meeting
  of the Association for Computational Linguistics (ACL'16)}, 2016.

\bibitem{nvidiadgx}
``The nvidia dgx-1 deep learning system.''
\newblock NVIDIA DGX-1.

\bibitem{Farabet09FPL}
C.~Farabet, C.~Poulet, J.~Y. Han, and Y.~LeCun, ``Cnp: An fpga-based processor
  for convolutional networks,'' in {\em Proceedings of the 2009 International
  Conference on Field Programmable Logic and Applications (FPL'09)},
  pp.~32--37, 2009.

\bibitem{Zhang15FPGA}
C.~Zhang, P.~Li, G.~Sun, Y.~Guan, B.~Xiao, and J.~Cong, ``Optimizing fpga-based
  accelerator design for deep convolutional neural networks,'' in {\em
  Proceedings of the 2015 ACM/SIGDA International Symposium on
  Field-Programmable Gate Arrays (FPGA'15)}, pp.~161--170, 2015.

\bibitem{Chen14ASPLOS}
T.~Chen, Z.~Du, N.~Sun, J.~Wang, and C.~Wu, ``{DianNao: a small-footprint
  high-throughput accelerator for ubiquitous machine-learning},'' in {\em
  Proceedings of the 19th international conference on Architectural support for
  programming languages and operating systems (ASPLOS)}, (Salt Lake City, UT,
  USA), pp.~269--284, 2014.

\bibitem{Farabet2011}
C.~Farabet, B.~Martini, B.~Corda, P.~Akselrod, E.~Culurciello, and Y.~LeCun,
  ``{NeuFlow: A runtime reconfigurable dataflow processor for vision},'' in
  {\em IEEE Computer Society Conference on Computer Vision and Pattern
  Recognition Workshops (CVPRW)}, pp.~109--116, Ieee, jun 2011.

\bibitem{Han2016a}
S.~Han, X.~Liu, H.~Mao, J.~Pu, A.~Pedram, M.~A. Horowitz, and W.~J. Dally,
  ``{EIE: Efficient Inference Engine on Compressed Deep Neural Network},'' in
  {\em Proceedings of the 43th Annual International Symposium on Computer
  Architecture (ISCA'16)}, vol.~16, 2016.

\bibitem{speccpu}
``The standard performance evaluation corporation (spec).''
\newblock SPEC CPU.

\bibitem{parsec}
C.~Bienia, S.~Kumar, J.~P. Singh, and K.~Li, ``{The PARSEC benchmark suite:
  Characterization and architectural implications},'' in {\em Proceedings of
  the International Conference on Parallel Architectures and Compilation
  Techniques}, pp.~72--81, 2008.

\bibitem{Alwani2016}
M.~Alwani, H.~Chen, M.~Ferdman, and P.~Milder, ``{Fused-layer CNN
  accelerators},'' in {\em 2016 49th Annual IEEE/ACM International Symposium on
  Microarchitecture (MICRO)}, pp.~1--12, 2016.

\bibitem{Judd2016}
P.~Judd, J.~Albericio, and A.~Moshovos, ``{Stripes: Bit-Serial Deep Neural
  Network Computing},'' in {\em 2016 49th Annual IEEE/ACM International
  Symposium on Microarchitecture (MICRO)}, vol.~6056, pp.~1--1, 2016.

\bibitem{Rhu2016}
M.~Rhu, N.~Gimelshein, J.~Clemons, A.~Zulfiqar, and S.~W. Keckler,
  ``{Virtualizing Deep Neural Networks for Memory-Efficient Neural Network
  Design},'' in {\em 2016 49th Annual IEEE/ACM International Symposium on
  Microarchitecture (MICRO)}, 2016.

\bibitem{Zhang2016a}
S.~Zhang, Z.~Du, L.~Zhang, H.~Lan, S.~Liu, L.~Li, Q.~Guo, T.~Chen, and Y.~Chen,
  ``{Cambricon-X : An Accelerator for Sparse Neural Networks},'' in {\em
  Proceedings of the 49th Annual IEEE/ACM International Symposium on
  Microarchitecture (MICRO-49)}, 2016.

\bibitem{Ji2016}
Y.~Ji, Y.~Zhang, S.~Li, P.~Chi, C.~Jiang, P.~Qu, Y.~Xie, and W.~Chen,
  ``{NEUTRAMS : Neural Network Transformation and Co-design under Neuromorphic
  Hardware Constraints},'' in {\em 2016 49th Annual IEEE/ACM International
  Symposium on Microarchitecture (MICRO)}, no.~October, 2016.

\bibitem{Kim2016a}
D.~Kim, J.~Kung, S.~Chai, S.~Yalamanchili, and S.~Mukhopadhyay, ``{Neurocube: A
  Programmable Digital Neuromorphic Architecture with High-Density 3D
  Memory},'' in {\em 2016 ACM/IEEE 43rd Annual International Symposium on
  Computer Architecture (ISCA)}, pp.~380--392, 2016.

\bibitem{LiKamWa2016}
R.~LiKamWa, Y.~Hou, Y.~Gao, M.~Polansky, and L.~Zhong, ``{RedEye: Analog
  ConvNet Image Sensor Architecture for Continuous Mobile Vision},'' in {\em
  2016 ACM/IEEE 43rd Annual International Symposium on Computer Architecture
  (ISCA)}, pp.~255--266, 2016.

\bibitem{Albericio2016}
J.~Albericio, P.~Judd, T.~Hetherington, T.~Aamodt, N.~E. Jerger, and
  A.~Moshovos, ``{Cnvlutin: Ineffectual-Neuron-Free Deep Neural Network
  Computing},'' in {\em 2016 ACM/IEEE 43rd Annual International Symposium on
  Computer Architecture (ISCA)}, pp.~1--13, 2016.

\bibitem{Chi2016}
P.~Chi, S.~Li, C.~Xu, T.~Zhang, J.~Zhao, Y.~Liu, Y.~Wang, and Y.~Xie, ``{PRIME:
  A Novel Processing-in-Memory Architecture for Neural Network Computation in
  ReRAM-Based Main Memory},'' in {\em 2016 ACM/IEEE 43rd Annual International
  Symposium on Computer Architecture (ISCA)}, pp.~27--39, 2016.

\bibitem{Shafiee2016}
A.~Shafiee, A.~Nag, N.~Muralimanohar, R.~Balasubramonian, J.~P. Strachan,
  M.~Hu, R.~S. Williams, and V.~Srikumar, ``Isaac: A convolutional neural
  network accelerator with in-situ analog arithmetic in crossbars,'' in {\em
  2016 ACM/IEEE 43rd Annual International Symposium on Computer Architecture
  (ISCA)}, pp.~14--26, June 2016.

\bibitem{Liu2016a}
S.~Liu, Z.~Du, J.~Tao, D.~Han, T.~Luo, Y.~Xie, Y.~Chen, and T.~Chen,
  ``{Cambricon: An Instruction Set Architecture for Neural Networks},'' in {\em
  2016 ACM/IEEE 43rd Annual International Symposium on Computer Architecture
  (ISCA)}, pp.~393--405, 2016.

\bibitem{Chen2016b}
Y.-H. Chen, J.~Emer, and V.~Sze, ``{Eyeriss: A Spatial Architecture for
  Energy-Efficient Dataflow for Convolutional Neural Networks},'' in {\em 2016
  ACM/IEEE 43rd Annual International Symposium on Computer Architecture
  (ISCA)}, pp.~367--379, 2016.

\bibitem{Reagen2016}
B.~Reagen, P.~Whatmough, R.~Adolf, S.~Rama, H.~Lee, S.~K. Lee, J.~M.
  Hernandez-Lobato, G.-Y. Wei, and D.~Brooks, ``{Minerva: Enabling Low-Power,
  Highly-Accurate Deep Neural Network Accelerators},'' in {\em 2016 ACM/IEEE
  43rd Annual International Symposium on Computer Architecture (ISCA)},
  pp.~267--278, 2016.

\bibitem{Song2017}
L.~Song, X.~Qian, H.~Li, and Y.~Chen, ``{PipeLayer : A Pipelined ReRAM-Based
  Accelerator for Deep Learning Basics of Deep Neural Network},'' in {\em
  Proceedings of The 23rd IEEE Symposium on High Performance Computer
  Architecture (HPCA)}, 2017.

\bibitem{Lu2017}
W.~Lu, G.~Yan, J.~Li, S.~Gong, Y.~Han, and X.~Li, ``{FlexFlow: A Flexible
  Dataflow Accelerator Architecture for Convolutional Neural Networks},'' in
  {\em Proceedings of The 23rd IEEE Symposium on High Performance Computer
  Architecture (HPCA)}, pp.~553--564, 2017.

\bibitem{Wzr2017}
W.~Wzr, V.~Surfhvv, L.~V. Ghsor, H.~G. Rq, K.~Hqg, D.~Rq, P.~D.~Q. Fkdoohqjlqj,
  P.~Ohduqlqj, H.~J. L.~W. Wdnhv, W.~Zhhnv, W.~R. Wudlq, R.~Q. Irxu, and
  K.~Hqg, ``{Towards Pervasive and User Staisfactory CNN across GPU
  Microarchitecture},'' in {\em Proceedings of The 23rd IEEE Symposium on High
  Performance Computer Architecture (HPCA)}, 2017.

\bibitem{Chen2012a}
T.~Chen, Y.~Chen, M.~Duranton, Q.~Guo, A.~Hashmi, M.~Lipasti, A.~Nere, S.~Qiu,
  M.~Sebag, and O.~Temam, ``{BenchNN: On the broad potential application scope
  of hardware neural network accelerators},'' {\em 2012 IEEE International
  Symposium on Workload Characterization (IISWC)}, pp.~36--45, nov 2012.

\bibitem{Baidu16DeepBench}
B.~Research, ``Deepbench,'' 2016.

\bibitem{Adolf2016}
R.~Adolf, S.~Rama, B.~Reagen, G.-y. Wei, and D.~Brooks, ``{Fathom: Reference
  Workloads for Modern Deep Learning Methods},'' in {\em 2016 IEEE
  International Symposium on Workload Characterization (IISWC)}, 2016.

\bibitem{Murtagh95JoC}
F.~Murtagh and M.~Hern{\'a}ndez-Pajares, ``The kohonen self-organizing map
  method: An assessment,'' {\em Journal of Classification}, vol.~12, no.~2,
  pp.~165--190, 1995.

\bibitem{Ren}
S.~Ren, K.~He, and R.~Girshick,
  ``{Faster-R-Cnn-Towards-Real-Time-Object-Detection-With-Region-Proposal-Networks},''
  in {\em Advances in neural information processing systems}, pp.~1--9, 2015.

\bibitem{Rastegari2016a}
M.~Rastegari, V.~Ordonez, J.~Redmon, and A.~Farhadi, ``{XNOR-Net: ImageNet
  Classification Using Binary Convolutional Neural Networks},'' {\em arXiv
  preprint}, pp.~1--17, 2016.

\bibitem{Song2015}
H.~Song, J.~Pool, J.~Tran, and W.~J. Dally, ``{Learning Both Weights and
  Connections for Efficient Neural Networks},'' in {\em Advances in Neural
  Information Processing Systems (NIPS'15)}, pp.~1135----1143, 2015.

\bibitem{YannLeCun1998}
Y.~LeCun, L.~L. Bottou, Y.~Bengio, and P.~Haffner, ``{Gradient-Based Learning
  Applied to Document Recognition},'' {\em Proceedings of the IEEE}, vol.~86,
  no.~11, pp.~2278-- 2324, 1998.

\bibitem{Graves2014}
A.~Graves and N.~Jaitly, ``{Towards End-To-End Speech Recognition with
  Recurrent Neural Networks},'' {\em JMLR Workshop and Conference Proceedings},
  vol.~32, no.~1, pp.~1764--1772, 2014.

\bibitem{Marcus1993}
M.~P. Marcus, B.~Santorini, and M.~A. Marcinkiewicz, ``{Building a large
  annotated corpus of English: The Penn Treebank.},'' {\em Computational
  Linguistics}, vol.~19, no.~2, pp.~313--330, 1993.

\bibitem{Russakovsky2014}
O.~Russakovsky, J.~Deng, H.~Su, J.~Krause, S.~Satheesh, S.~Ma, Z.~Huang,
  A.~Karpathy, A.~Khosla, M.~Bernstein, A.~C. Berg, and L.~Fei-Fei, ``{ImageNet
  Large Scale Visual Recognition Challenge},'' p.~37, sep 2014.

\bibitem{He}
K.~He, ``{Deep Residual Learning for Image Recognition},''

\bibitem{pascal-voc-2012}
M.~Everingham, L.~Van~Gool, C.~K.~I. Williams, J.~Winn, and A.~Zisserman, ``The
  {PASCAL} {V}isual {O}bject {C}lasses {C}hallenge 2012 {(VOC2012)}
  {R}esults.''
  http://www.pascal-network.org/challenges/VOC/voc2012/
  workshop/index.html.

\bibitem{Parkhi2015}
O.~M. Parkhi, A.~Vedaldi, and A.~Zisserman, ``{Deep Face Recognition},'' in
  {\em Procedings of the British Machine Vision Conference 2015}, no.~Section
  3, pp.~41.1--41.12, 2015.

\bibitem{LFWTech}
G.~B. Huang, M.~Ramesh, T.~Berg, and E.~Learned-Miller, ``Labeled faces in the
  wild: A database for studying face recognition in unconstrained
  environments,'' Tech. Rep. 07-49, University of Massachusetts, Amherst,
  October 2007.

\bibitem{Noh2015}
H.~Noh, S.~Hong, and B.~Han, ``{Learning Deconvolution Network for Semantic
  Segmentation},'' in {\em The IEEE International Conference on Computer Vision
  (ICCV)}, vol.~1, 2015.

\bibitem{Johnson2015}
J.~Johnson, A.~Karpathy, and L.~Fei-Fei, ``{DenseCap: Fully Convolutional
  Localization Networks for Dense Captioning},'' {\em arXiv preprint}, 2015.

\bibitem{Venugopalan}
S.~Venugopalan, M.~Rohrbach, T.~Darrell, J.~Donahue, K.~Saenko, and R.~Mooney,
  ``{Sequence to Sequence - Video to Text},'' in {\em Proceedings of the IEEE
  International Conference on Computer Vision}, pp.~4534--4542, 2015.

\bibitem{Chen2011}
D.~L. Chen and W.~B. Dolan, ``{Collecting Highly Parallel Data for Paraphrase
  Evaluation},'' {\em Proceedings of the 49th Annual Meeting of the Association
  for Computational Linguistics}, pp.~190--200, 2011.

\bibitem{Andor2016}
D.~Andor, C.~Alberti, D.~Weiss, A.~Severyn, A.~Presta, K.~Ganchev, S.~Petrov,
  and M.~Collins, ``{Globally Normalized Transition-Based Neural Networks},''
  in {\em arXiv preprint}, pp.~2442--2452, 2016.

\bibitem{Phansalkar07ISCA}
A.~Phansalkar, A.~Joshi, and L.~K. John, ``Analysis of redundancy and
  application balance in the spec cpu2006 benchmark suite,'' in {\em
  Proceedings of the 34th Annual International Symposium on Computer
  Architecture (ISCA'07)}, pp.~412--423, 2007.

\bibitem{McCalpin95TCCA}
J.~D. McCalpin, ``Memory bandwidth and machine balance in current high
  performance computers,'' {\em IEEE Computer Society Technical Committee on
  Computer Architecture (TCCA) Newsletter}, pp.~19--25, 1995.

\bibitem{Bull01CAN}
J.~M. Bull and D.~O'Neill, ``A microbenchmark suite for openmp 2.0,'' {\em
  SIGARCH Comput. Archit. News}, vol.~29, no.~5, pp.~41--48, 2001.

\bibitem{Du2015}
Z.~Du, R.~Fasthuber, T.~Chen, P.~Ienne, L.~Li, X.~Feng, Y.~Chen, and O.~Temam,
  ``{ShiDianNao: Shifting Vision Processing Closer to the Sensor},'' in {\em
  Proceedings of the 42nd Annual International Symposium on Computer
  Architecture}, pp.~92--104, 2015.

\bibitem{Ding03PLDI}
C.~Ding and Y.~Zhong, ``Predicting whole-program locality through reuse
  distance analysis,'' in {\em Proceedings of the ACM SIGPLAN Conference on
  Programming Language Design and Implementation (PLDI)}, pp.~245--257, 2003.

\bibitem{Chen2015}
Y.~Chen, T.~Luo, S.~Liu, S.~Zhang, L.~He, J.~Wang, L.~Li, T.~Chen, Z.~Xu,
  N.~Sun, and O.~Temam, ``{DaDianNao: A Machine-Learning Supercomputer},'' in
  {\em Proceedings of the 47th Annual IEEE/ACM International Symposium on
  Microarchitecture (MICRO-47)}, pp.~609--622, 2015.

\bibitem{Pawlowski11Hotchips}
J.~T. Pawlowski, ``Hybrid memory cube: breakthrough dram performance with a
  fundamentally re-architected dram subsystem,'' in {\em Proceedings of the
  23rd Hot Chips Symposium (HotChips'11)}, 2011.

\bibitem{Jia14Caffe}
Y.~Jia, E.~Shelhamer, J.~Donahue, S.~Karayev, J.~Long, R.~Girshick,
  S.~Guadarrama, and T.~Darrell, ``Caffe: Convolutional architecture for fast
  feature embedding,'' {\em arXiv preprint arXiv:1408.5093}, 2014.

\bibitem{Denkowski2014}
M.~Denkowski and A.~Lavie, ``{Meteor Universal: Language Specific Translation
  Evaluation for Any Target Language},'' pp.~376--380, 2014.

\bibitem{Mucci99PAPI}
P.~J. Mucci, S.~Browne, C.~Deane, and G.~Ho, ``Papi: A portable interface to
  hardware performance counters,'' in {\em In Proceedings of the Department of
  Defense HPCMP Users Group Conference}, pp.~7--10, 1999.

\bibitem{maxwell}
{NVIDIA}, ``Nvidia tegra x1: Nvidia's new mobile superchip.''

\bibitem{950}
{NVIDIA}, ``Geforce gtx 950m: specifications.''

\bibitem{k40}
{NVIDIA}, ``Nvidia tesla gpu accelerators.''

\bibitem{Keckler2011}
S.~W. Keckler, W.~J. Dally, B.~Khailany, M.~Garland, and D.~Glasco, ``{GPUs and
  the future of parallel computing},'' {\em IEEE Micro}, vol.~31, no.~5,
  pp.~7--17, 2011.

\bibitem{graves2013speech}
A.~Graves, A.-r. Mohamed, and G.~Hinton, ``Speech recognition with deep
  recurrent neural networks,'' in {\em 2013 IEEE international conference on
  acoustics, speech and signal processing}, pp.~6645--6649, IEEE, 2013.

\bibitem{Karpathy2015}
A.~Karpathy and F.~Li, ``Deep visual-semantic alignments for generating image
  descriptions,'' in {\em {IEEE} Conference on Computer Vision and Pattern
  Recognition, {CVPR} 2015, Boston, MA, USA, June 7-12, 2015}, pp.~3128--3137,
  2015.

\bibitem{He2015}
K.~He, X.~Zhang, S.~Ren, and J.~Sun, ``Delving deep into rectifiers: Surpassing
  human-level performance on imagenet classification,'' in {\em Proceedings of
  the 2015 IEEE International Conference on Computer Vision (ICCV)}, ICCV '15,
  (Washington, DC, USA), pp.~1026--1034, IEEE Computer Society, 2015.

\bibitem{Taigman2014}
Y.~Taigman, M.~Yang, M.~Ranzato, and L.~Wolf, ``Deepface: Closing the gap to
  human-level performance in face verification,'' in {\em Proceedings of the
  2014 IEEE Conference on Computer Vision and Pattern Recognition}, CVPR '14,
  (Washington, DC, USA), pp.~1701--1708, IEEE Computer Society, 2014.

\bibitem{Le2012}
Q.~V. Le, M.~Ranzato, R.~Monga, M.~Devin, G.~Corrado, K.~Chen, J.~Dean, and
  A.~Y. Ng, ``Building high-level features using large scale unsupervised
  learning,'' in {\em Proceedings of the 29th International Conference on
  Machine Learning, {ICML} 2012, Edinburgh, Scotland, UK, June 26 - July 1,
  2012}, 2012.

\end{thebibliography}
\end{document}